\documentclass[journal]{IEEEtran} 

\IEEEoverridecommandlockouts 

\usepackage[noadjust]{cite}
\usepackage{mathtools, lipsum}
\usepackage{amsmath,amssymb,mathrsfs,amsfonts,dsfont,amsthm,soul,yhmath,accents}
\usepackage{enumerate}
\usepackage[english]{babel}
\usepackage{comment}
\usepackage{epsfig,amsfonts,latexsym,amssymb,shadow,amsmath,wrapfig}
\usepackage[usenames]{color}
\usepackage{algorithm}
\usepackage{algpseudocode}
\usepackage{algorithm, accents}
\usepackage{footnote}
\usepackage[dvipsnames]{xcolor}
\usepackage{tablefootnote}
\makesavenoteenv{tabular}

\newtheorem{theorem}{Theorem}

\newtheorem{lemma}{Lemma}
\newtheorem{prop}{Proposition}
\newtheorem{remark}{Remark}
\newtheorem{defn}{Definition}

\newtheorem{assum}{Assumption}

\DeclareMathOperator*{\argmax}{\arg\!\max}

\newcommand{\ubar}[1]{\underaccent{\bar}{#1}}

\newcounter{l1}
\newcounter{l2}
\newcounter{l3}
\newcommand{\bdotlist}{\begin{list}{$\bullet$}{}}
\newcommand{\bboxlist}{\begin{list}{$\Box$}{}}
\newcommand{\bbboxlist}{\begin{list}{\raisebox{.005in}{{\tiny $\blacksquare$ \ \ }}}{}}
\newcommand{\bdashlist}{\begin{list}{$-$}{} }
\newcommand{\blist}{\begin{list}{}{} }
\newcommand{\barablist}{\begin{list}{\arabic{l1}}{\usecounter{l1}}}
\newcommand{\balphlist}{\begin{list}{(\alph{l2})}{\usecounter{l2}}}
\newcommand{\bAlphlist}{\begin{list}{\Alph{l2}.}{\usecounter{l2}}}
\newcommand{\bdiamlist}{\begin{list}{$\diamond$}{}}
\newcommand{\bromalist}{\begin{list}{(\roman{l3})}{\usecounter{l3}}}

\begin{document}
%
\title{Optimal Dynamic Mechanism Design with Stochastic Supply and Flexible Consumers}
%
%
%

\author{Shiva~Navabi,~\IEEEmembership{Student Member,~IEEE,}
        Ashutosh~Nayyar,~\IEEEmembership{Member,~IEEE}
\thanks{S. Navabi and A. Nayyar are with the Electrical and Computer Engineering Department, University of Southern California, 3740 McClintock Avenue, Los Angeles, CA, 90089 USA (E-mail: navabiso@usc.edu; ashutosn@usc.edu.)}
}

\maketitle

\begin{abstract}

We consider the problem of designing an expected-revenue maximizing mechanism for allocating multiple non-perishable goods of $k$ varieties  to flexible consumers over $T$ time steps. In our model, a random number of goods of each variety may become available to the seller at each time and a random number of consumers may enter the market at each time. Each consumer is present in the market for one time step and wants to consume one good of one of its desired varieties. Each consumer is associated with a flexibility level that indicates the varieties of the goods it is equally interested in. A consumer's flexibility level and the utility it gets from consuming a good of its desired varieties are its private information. We characterize the allocation rule for a Bayesian incentive compatible, individually rational and expected revenue maximizing mechanism in terms of the solution to a dynamic  program. The corresponding payment function is also  specified in terms of  the optimal allocation function. We leverage the  structure of the consumers' flexibility model to simplify  the dynamic program and provide an alternative description of the optimal mechanism in terms of thresholds computed by the dynamic program.


\end{abstract}


%
\IEEEpeerreviewmaketitle

\section{Introduction}\label{sec:intro}
Consider the scenario faced by a monopolist seller with multiple resources who wants to allocate them to consumers over time in order to maximize its expected total revenue. The seller offers goods of different varieties and there may be new additions to its stock of each variety over time.     Different consumers may interact  with the market  at various points in time, each for a limited duration. Such a scenario arises in many marketplaces where the available supply and  the population of the consumers  vary in an uncertain fashion over time. In cloud computing platforms \cite{agmon2013deconstructing}, for example, the computational and data storage resources get  freed up with the termination of  previously submitted jobs  dynamically over time  and are dedicated to processing of the new tasks   as they are received by the platform. In power distribution networks with partial reliance on the renewable energy resources \cite{rogers2012delivering} the energy supply varies over time depending on the availability of the intermittent source of  energy; the amount of   power demanded by the consumer base  connected to the grid  is also uncertain and constantly fluctuates over time. In wireless spectrum management platforms \cite{hossain2009dynamic} the available spectrum bands are leased to secondary wireless service providers for temporary usage  and are freed up as the interim lease contracts terminate dynamically over time. One particular feature that makes these resource allocation problems  challenging is that in the face of uncertainty about the future supply and demand  the seller needs to decide whether to use its limited resources to serve a  currently present consumer or  keep them for   potentially more profitable  transactions in  future.  Moreover, in order to decide about the optimal way of allocating its resources, the seller needs the information about the consumers' preferences and restrictions,  and their willingness to pay for their desired goods or services. Typically, however, this information is   known privately to each consumer and the seller needs to \textit{elicit} this data from them.    Since the consumers are self-interested and strategic, they may distort their privately held information  when communicating it to the seller if they believe that they can benefit from such misrepresentations. The seller thus needs to provide \textit{incentives} to the consumers in  a way that they would find it in their own best interest to disclose their private information \textit{truthfully} to the seller. The theory of mechanism design provides a systematic framework for designing the rules of interaction between multiple strategic   agents in  a way  that  the principal decision maker's desired outcome emerges  at the equilibrium of the induced game.

Auctions as a special class of mechanisms  have been extensively studied in the  context of mechanism design. Due to their practical advantages over conventional selling schemes such as menu-based pricing,  auctions are being  adopted in an increasing number of  marketplaces  for pricing and selling various products and services.   
While the theory of auction design is well developed under static settings, its extension to dynamic markets that involve allocation and pricing  of time-varying  supply to accommodate time-varying demand under incomplete information is generally less mature and is still an active area of research \cite{bergemann2019dynamic}.
Given the  growing  practical interest in exploring the potential of auction mechanisms for allocating and pricing resources  in dynamically operating markets, a deeper understanding of the design and implementation of such auctions is crucial. 

In this paper we study the problem of designing expected-revenue maximizing auctions for  selling  indivisible and durable goods of $k$  varieties to consumers   over a discrete, finite time horizon. Additional units of each variety may become available to the seller at each time step.   In our setup, each consumer is associated with a flexibility level which indicates the varieties of the goods that the consumer  finds equally  desirable. Formally, the flexibility level of a consumer is a number in the range $1, \ldots, k$ such that a consumer with flexibility  level $j$ wants to get a good of any of the first $j$ varieties.  Each consumer is present in the market for one time period and  wants to receive one good  (of any of the desired varieties) prior to  its departure. The flexibility level and the valuation a consumer has for a  desirable good are both its private information. 

There are several markets with a temporally fluctuating consumer base where the flexibility in demand described above arises. We describe two such scenarios below.\\

\textit{1. Dynamic Spectrum Management in Cognitive Radio Networks:}
With the emergence of various wireless applications for mobile users, there has been significant increase in the demand for radio frequency spectrum in recent years. While most of the available radio spectrum has already been licensed off to the existing wireless service providers (WSPs), they are not being fully utilized by their primary owners. As pointed out in \cite{hossain2009dynamic},   dynamic spectrum access protocols can enable efficient use of these underutilized frequency bands -- referred to as \textit{spectrum holes} in \cite{hossain2009dynamic} -- by accommodating the demands of secondary users who can use these bands on a temporary basis.
Cognitive radio systems can detect the presence of such spectrum holes in the frequency bands owned by a primary user.
Consider the problem faced by  one such spectrum owner who wants to allocate its underutilized frequency bands  of various widths, as they become available over time, to secondary WSPs who have different minimum bandwidth requirements. Suppose the primary owner has frequency bands of widths $w_1, \ldots, w_k$ such that $w_1 > w_2 > \cdots > w_k$. We say that a WSP is  of flexibility level $j$ if it requires  a frequency band of width at least $w_{j}$. At the beginning of each time step $t$ a random number of  WSPs  arrive into the market to compete for the limited radio frequency bands available at that time. The  resource allocation problem that a primary spectrum owner encounters when it aims to allocate its intermittently available frequency bands to secondary users' temporary usage can be modeled within the framework developed in the present paper.\\

\textit{2. Allocation of Computational Resources in Cloud Computing Platforms:} 
Consider Amazon's EC2 cloud computing platform that sells  various types of computational resources such as memory, CPU, storage cpacity, computer applications, etc. In this market, clients can randomly enter and depart over time. Clients rent virtual machines or \textit{instances} and are typically charged on an hourly basis per instance depending on the duration of their usage as well as the rented instance type.  Each of the instance types are offered in different sizes to suit various purposes. As explained in \cite{agmon2013deconstructing}, for example, Amazon's EC2 offers ``standard" instances  in three sizes: small, large and extra-large.
A consumer can belong to one of three flexibility  classes:
\begin{itemize}
\item \textit{Inflexible consumers} who need an extra-large instance.
\item \textit{Somewhat flexible consumers} who need a large or extra-large instance.
\item \textit{Flexible consumers} who are fine with receiving any of the three types of instances.
\end{itemize}
The allocation of computing instances to consumers of differing flexibilities can be modeled by our setup.

\subsection{Prior Work}
Much of the prior work in the area of dynamic auctions can be grouped in two categories \cite{bergemann2010dynamic}:
1)  markets with a dynamic  population of consumers whose private information remain unchanged over time, and 2) markets with a fixed population of consumers whose private information evolve over time. Within each of these two categories the important findings on efficiency (social-welfare maximization) and optimality (revenue maximization) as the two primary objectives are highlighted in  \cite{bergemann2010dynamic}. Our work falls under the first category (dynamic population) above with the focus on revenue maximization as the main objective. 
Therefore, we will focus on comparing our model with prior works that have addressed revenue maximization under the first category above.
We can broadly categorize the works in this strand of the literature based on  certain features of the seller's supply and the consumers' demands as follows:

\subsubsection{Dynamic Auctions with Multiple Identical and Durable Goods}\label{sec:identDur}

The works in this area have studied dynamic revenue-maximizing  auctions in settings where the seller has multiple identical goods and wants to sell them to unit-demand\footnote{A unit-demand consumer wants to receive one unit of the good \cite{nisan2007algorithmic}.} consumers over a finite or infinite time horizon.  Vulcano et al \cite{vulcano2002optimal} consider  a setup where the seller has an initial capacity of $C$ units of the same good and wants to sell them to unit-demand consumers over $T$ time steps where each consumer is present in the market for one time period. Pai and Vohra \cite{pai2013optimal} consider the same setup  under the assumption that each consumer can have a presence time window longer than one time period where each consumer's entry and exit times are its private information.  Lavi and Nisan \cite{lavi2004competitive} consider a  setup similar to the one in \cite{pai2013optimal} where each consumer may demand more than one unit of the good.    Gallien \cite{gallien2006dynamic} studies a similar setup where the seller  offers $K$ identical items for sale over an infinite  time horizon and  consumers are assumed to be unit-demand and time sensitive in the sense that they discount their future utility with a common time-discount factor.   Gershkov et al \cite{gershkov2017revenue} design a revenue-maximizing mechanism for a setting where the seller has multiple identical goods for sale over an infinite continuous time horizon. They assume in \cite{gershkov2017revenue} that the  consumers are unit-demand and  that each consumer's arrival time as well as its valuation are its private information. 



The key feature  that differentiates  these setups from our model   is that in all of them  the  goods  are assumed to be \textit{identical}. In our model, each consumer, depending on its flexibility level,  \textit{subjectively differentiates} between the goods. In particular, a consumer with flexibility level $j$ has the same positive valuation for any good of varieties $1, \ldots, j$ and zero valuation for a good of  varieties $j+1, \ldots, k$.
Furthermore, in  our model, more units may be added to the seller's supply of different varieties  over time. In the setups described above,   the seller's supply is limited to the initial stock of the goods available at the beginning of the time horizon.

\subsubsection{Dynamic Auctions with Multiple Identical and Perishable Goods} 
Said \cite{said2012auctions} considers a setup where a seller obtains  an uncertain number of perishable\footnote{A good is perishable if it cannot be stored for future allocations.} identical goods  at each time step and wants to sell them to unit-demand consumers over an infinite discrete time horizon. Each consumer may depart the market exogenously at any time period $t$ after its arrival with a common probability $(1-\gamma_t) \in [0,1]$. Otherwise, a consumer continues to interact with the market until it gets an allocation.  Unlike the setup in \cite{said2012auctions}, in our model we assume that goods are durable and could thus be stored for future allocations. Moreover, in contrast to \cite{said2012auctions}, we assume that consumers are present in the market for one time step only. Lastly,  in our model goods are valued differently by different consumers  depending on their flexibility levels whereas in \cite{said2012auctions}  the offered goods are all identical from the consumers' viewpoints.

\subsubsection{Dynamic Auctions with Multiple Heterogeneous Goods} 
The papers in this line of work study  dynamic revenue-maximizing auction design problem in cases where the seller has multiple \textit{heterogeneous} goods and wants to sell them to unit-demand consumers over a finite/infinite time horizon. Gershkov and Moldovanu in \cite{gershkov2009dynamic} study one such setup over a continuous and finite time horizon where the goods are \textit{commonly ranked} by the consumers  that are impatient, i.e., they want to get an allocation immediately upon arrival in the market. Unlike this setup, each consumer in our model  has \textit{subjective} preferences for different varieties of the goods    and each consumer's desired varieties are its private information. 
Furthermore, in our model, additional goods of each variety may be added to the seller's supply over time which is not the case in \cite{gershkov2009dynamic}. 

\subsubsection{Dynamic Auctions with Private  Departure Times}  
Mierendorff \cite{mierendorff2016optimal} considers a setup where a seller wants to sell a single indivisible good over $T$ time steps to consumers who are privately informed about their valuations as well as  their deadlines for buying the single item in the dynamic auction.  The key differences between this work and our setup are: 1) in our model the seller offers multiple goods that are differentiated by each consumer subjectively based on their privately known flexibility levels while in \cite{mierendorff2016optimal} the seller has only one good for sale, 2) the consumers' exit times are known to the seller in our setup while they are  privately known to each consumer  in \cite{mierendorff2016optimal} and, 3)  in our model each consumer is present in the market for one time step only  while in \cite{mierendorff2016optimal} a consumer may remain present in the market for multiple time periods.

In the model studied by Pai  and Vohra in \cite{pai2013optimal}   each consumer's departure time  is assumed to be its private information. As mentioned in Section \ref{sec:identDur} however, unlike our model, in the setup in \cite{pai2013optimal} goods are assumed to be identical.

\subsection{Notations}
Random variables are denoted by upper case letters $(X, Y, N)$ or by Greek letters $(\theta)$, their realizations by the corresponding lower case letters $(x, y, n)$, or by   Greek letters with tilde $(\tilde{\theta})$. $\{ 0,1\}^{N \times M}$ denotes the space of $N\times M$  matrices with entries that are either 0 or 1. $\bold{0}_{1\times k}$ is the $k$-dimensional all-zeros row vector. $\mathbb{Z}_{\ge 0}$ and $\mathbb{Z}_{+}$ denote the sets of non-negative and positive integers, respectively. For a set $\mathcal{A}$, $|\mathcal{A}|$ denotes the cardinality of $\mathcal{A}$.  $x^+$ is the positive part of the real number $x$, that is, $x^+ = \max(x,0)$. 
$\mathds{1}_{\{a\le b\}}$ denotes 1 if the inequality in the subscript is true and 0 otherwise. $\mathbb{E}$ denotes the expectation operator. For a random variable/random vector $\theta$, $\mathbb{E}_{\theta}$ denotes that the expectation is with respect to the probability distribution of $\theta$.\footnote{The subscript for $\mathbb{E}[\cdot]$ operator is added only when its absence is likely to cause ambiguity.}  $x_{1:n}$, $y^{1:m}$ and $z^{1:m}_{1:n}$ are  shorthands for vectors $(x_1, \ldots, x_n)$, $(y^1, \ldots, y^m)$ and $(z_1^1, \ldots,z_1^m, \ldots, z_n^1, \ldots,z_n^m)$, respectively. For the vector $y^{1:m}$, $y^{-j}$ is the shorthand for $(y^1, \ldots, y^{j-1}, y^{j+1}, \ldots, y^m)$. The summation $\sum\limits_{i=m}^n y_i$ equals zero when $n < m$ regardless of  the indexed quantities $y_i$. 

\section{Problem Formulation}\label{sec:Formul}
We consider a  setup where a seller offers $k$ varieties of goods for sale over $T$ time steps. 
At  each time step a random number of consumers enter the market. Let the random variable $N_t$ denote the number of consumers that  arrive at  time step $t$. $N_t$ is an     integer-valued random variable that takes values in the set $\{0, 1, \ldots, \bar{n}\}$ according to the probability mass function (PMF) $\lambda_t(\cdot)$. 
At  each time step a random number of goods of varieties $1, 2, \ldots, k$  become available to the seller. Let the random variable $X_t^j$ denote the  number of goods of variety $j$ that become available at  time step $t$. $X_t^j$ is an integer-valued random variable that takes values in the set $\{ 0, 1, \ldots, \bar{x}^j \}$ according to the PMF $\gamma_t^j(\cdot)$. The random variables $N_{1:T}, X_{1:T}^{1:k}$, are mutually independent. Let $Y^j_t$ denote the number of unallocated goods of variety $j$ at time $t$ --- this includes $X^j_t$ as well as any unallocated variety $j$ goods from the past.  Let $V_t^j$ denote the  number of variety $j$ goods allocated by the seller at time $t$. $Y^j_t$  evolves according to the following dynamics:
\begin{align} 
Y_{t+1}^j &= \sum\limits_{s=1}^{t+1} X_s^j - \sum\limits_{s=1}^{t} V_s^j = Y_{t}^j - V_t^j + X_{t+1}^j, \; t \ge 1, \notag \\
Y^j_1 &= X_1^j \; , \; j = 1, 2, \ldots, k. \label{eq:Yjt}
\end{align}
\subsection{Consumer Flexibility and Consumer Type} \label{sec:flex}
Each consumer can consume at most one good.  Each consumer  has a flexibility level that indicates the varieties of goods the consumer is equally interested in. A consumer with flexibility level $1$ wants to get one good of variety $1$, a consumer with flexibility level $2$ wants to get one good of  either variety $1$ or variety $2$, and in general, a consumer with flexibility level $j$ wants to get one good of any of the first $j$ varieties.   

Each consumer is associated with a 4-tuple $(\theta, b, t^A, t^D)$ where:
\begin{enumerate}
\item $\theta$ is the consumer's utility if it receives one good of a desired variety. We refer to $\theta$ as the consumer's valuation.
\item $b$ is the consumer's flexibility level.
\item $t^A$ is the consumer's arrival time.
\item $t^D$ is the consumer's departure time.
\end{enumerate}
A consumer can receive a good at any time $t$, $t^A \le t \le t^D$.
\begin{defn}
We say that a consumer is impatient \cite[Chapter 16]{nisan2007algorithmic} if its arrival and departure times are the same. Thus, an impatient consumer can only receive a good at its arrival time.
\end{defn}
In this paper we assume that all consumers are impatient. 

The random variable $b_t^i$ denotes the flexibility level of the $i$th consumer  arriving at time $t$. $b^i_t$ takes values in the set $\{1, \ldots, k\}$ according to the PMF $g_t(\cdot)$.  
The random variable $\theta_t^i$ denotes the valuation of the $i$th consumer arriving at time $t$. Given $b^i_t = j$, $\theta_t^i$ takes values in $\Theta := [\theta^{\text{min}}, \theta^{\text{max}}]$ with conditional probability  density $\pi_t(\cdot | b^i_t=j)$. We define the joint distribution function  $f_t(\tilde{\theta}, j) := \pi_t(\tilde{\theta} | j) . g_t(j) \; , \; j \in \{ 1, \ldots, k\}, \tilde{\theta} \in \Theta$. The probability distributions $\lambda_t(\cdot), f_t(\cdot), \gamma_t^j(\cdot), \forall t, \forall j$ are common knowledge. 

For a consumer with valuation $\tilde{\theta}$ and flexibility level $\tilde{b}$ we refer to the pair $(\tilde{\theta}, \tilde{b})$ as its \textit{type}. Each consumer's type is independent of the other consumers' types and of the random variables $N_{1:T}, X_{1:T}^{1:k}$.

\subsection{Direct Mechanisms} \label{sec:direct}
We consider a direct mechanism where each consumer arriving in the market reports a valuation from the set $\Theta$ and a flexibility level from the set $\{ 1, 2, \ldots, k \}$. 
Each consumer can misreport its valuation and flexibility level. Consider a consumer whose true type is $(\tilde{\theta},\tilde{b})$ and let $(r,c)$ denote the type it reports, where $r$ is the reported valuation and $c$ is the reported flexibility level.  The consumers' arrivals  are publicly  observed.  Hence, $N_t$ is observed by the seller at time $t$ and by the consumers who arrive at time $t$. 
We make the following assumptions about the consumers' reported types:
\begin{assum}\label{assum:typeAs}
We assume that:
\begin{enumerate}
\item Each consumer reports its valuation and flexibility level simultaneously  at its arrival time.
\item No consumer departs the market without reporting a type to the seller.
\item Consumers  cannot over-report their flexibility levels, that is, $c$ cannot exceed $\tilde{b}$.
\end{enumerate}
\end{assum}

\subsection{Feasible Allocations}\label{sec:feasA}
Suppose $n_t$ consumers arrive at time  $t$, i.e., $N_t = n_t$.  Let $h^R_{t} := \{ (r^1_t, c^1_t), \ldots, (r_t^{n_t}, c^{n_t}_t) \}$ be the collection of reports made by the consumers arriving at time $t$,  where $(r_t^i, c_t^i)$ denotes the  type reported by the $i$th consumer arriving at time $t$. If $n_t = 0$, then $h^R_t = \emptyset$. Let $\mathcal{H}^R_t$ denote the set of all possible values of  $h^R_t$.
At each time $t$ if $h^R_t \neq \emptyset$, an allocation of the available goods among the currently present consumers  can be described by a  binary matrix $\bold{A}_t \in \{0,1\}^{n_{t}\times k}$.  $\bold{A}_t(i,j) = 1$ if the $i$th consumer is allocated a good of the $j$th variety  at time $t$ and $\bold{A}_t(i,j) = 0$ otherwise. The matrix $\bold{A}_t$ is called an allocation matrix at time $t$. 
$\bold{A}_t$ must satisfy some feasibility constraints. In particular, $\sum\limits_{i=1}^{n_t} \bold{A}_t(i,j) \le y^j_t, \forall j$, where $y^j_t$ is the number of variety $j$ goods available for allocation at time $t$. Further, we require that each consumer is allocated at most one good of its desired varieties and no goods of its undesired varieties, i.e., $\sum\limits_{j \le c^i_t} \bold{A}_t(i,j) \le 1, \sum\limits_{j > c^i_t} \bold{A}_t(i,j) = 0$ for $i = 1, \ldots, n_t$.
A binary matrix  that satisfies these  constraints is called a \textit{feasible} allocation matrix at time $t$. For $h^R_t \neq \emptyset$, let $\mathcal{S}(h^R_{t}, y_t^{1:k}) \subset \{0,1\}^{n_t\times k}$ denote the set of all feasible allocation matrices at time $t$. 
\subsection{Mechanism Setup}\label{sec:MechSetup}
Let $h_t$ denote all  the information that the seller knows  at time $t$. We call $h_t$ the history at time $t$ which is given as:
\begin{align} \label{eq:ht}
h_t \coloneqq \Big\{ h^R_{1:t} \; , \; y_{1:t}^{1:k} \; , \; x_{1:t}^{1:k} \Big\}.
\end{align}  
Let $\mathcal{H}_t$ denote the set of all possible values of $h_t$. We use $H_t$ to denote a random history.

A mechanism needs to specify  allocations and payments at each time $t$ for which the number of arriving consumers is nonzero, i.e., $h^R_t \neq \emptyset$. Such a   mechanism consists of the following components:
\begin{itemize}
\item A sequence of allocation functions $q_{1:T}$ such that for any $h_t$ with $h^R_t \neq \emptyset$, $q_t(h_t) \in \mathcal{S}(h^R_{t}, y_t^{1:k})$. $q_t(h_t)$ describes the allocation matrix to be used at time $t$.
\item A  sequence of payment functions $p_{1:T}$ such that for any $h_t$ with $h^R_t \neq \emptyset$, $p_t(h_t) \in \mathbb{R}^{|h^R_t|}$. The $i$th component $p^i_t(h_t)$ of $p_t(h_t)$ describes the payment charged to the $i$th consumer at time $t$.
\end{itemize}

\subsection{Consumer Utility Model}\label{sec:util}
Suppose $h_t$ is the history at time $t$ and $(\tilde{\theta}^i_t, \tilde{b}^i_t)$ is the true type of the $i$th consumer arriving at time $t$. Then, under the mechanism $(q_{1:T}, p_{1:T})$, this consumer's utility is given as
\begin{align}
u(\tilde{\theta}^i_t, \tilde{b}^i_t, h_{t}) = \tilde{\theta}^i_t \: \Big( \sum\limits_{j \le \tilde{b}^i_t} q^{i,j}_{t}(h_{t}) \Big) - p_{t}^{i}(h_{t}),  
\label{eq:utility}
\end{align}
where $q^{i,j}_{t}(h_{t})$ is the entry in the $i$th row and $j$th column of the allocation matrix $q_{t}(h_{t})$ and $p_{t}^{i}(h_{t})$ is the $i$th entry of the payments vector $p_{t}(h_{t})$. 

\subsection{Incentive Compatibility and Individual Rationality} \label{sec:BIC-IR}
The seller needs to design a mechanism that satisfies incentive compatibility and individual rationality constraints as described below.

In a Bayesian incentive compatible (BIC) mechanism, truthful reporting of private information (valuations and flexibility levels in our setup) constitutes an equilibrium of the Bayesian game induced by the mechanism. In other words, each consumer would prefer to report its true type  provided that all other consumers have adopted truth-telling strategy. Moreover, according to the revelation principle \cite{borgers2015introduction}, restriction to incentive compatible direct mechanisms is without loss of generality.

Suppose $n_t$ consumers arrive at time $t$ and let $(\tilde{\theta}^i_t, \tilde{b}^i_t)$ be the true type of the $i$th consumer arriving at time $t$. Recall that $N_t$ is observed by the consumers who arrive at time $t$ (see Section \ref{sec:direct}). Assuming that all other consumers report their types truthfully,  consumer $i$'s expected utility if  it reports its type truthfully will be
\begin{equation}\label{truthEU}
\mathbb{E}_{H_{t}^{-i}}\Big[ \; \tilde{\theta}^i_t \:  \sum\limits_{j \le \tilde{b}^i_t} q^{i,j}_{t}(H^{-i}_{t}, (\tilde{\theta}^i_t, \tilde{b}^i_t))  - p_{t}^{i}(H_t^{-i}, (\tilde{\theta}^i_t, \tilde{b}^i_t)) \;  | \; N_t = n_t \; \Big],
\end{equation}
where the expectation is with respect to the collection of random variables $H^{-i}_t$ which includes all variables in the history at time $t$ except the $i$th consumer's report.


Now suppose the consumer with type $(\tilde{\theta}^i_t, \tilde{b}^i_t)$ reports $(r^i_t, c^i_t)$ as its type rather than  $(\tilde{\theta}^i_t, \tilde{b}^i_t)$. That is, the consumer might  \textit{misreport} its valuation or its flexibility level or both. Assuming that all other consumers  truthfully  report their types, this consumer's expected utility if it reports $(r^i_t,c^i_t)$ will be
\begin{equation}\label{etaEU}
\mathbb{E}_{H_t^{-i}}\Big[ \; \tilde{\theta}^i_t \:  \sum\limits_{j \le \tilde{b}^i_t} q^{i,j}_{t}(H_{t}^{-i}, (r^i_t, c^i_t))  - p_{t}^{i}(H_{t}^{-i}, (r^i_t, c^i_t)) \;  | \; N_t = n_t \; \Big].
\end{equation}
BIC constraint is satisfied if  each consumer's  expected utility is maximized when it reports its valuation and flexibility level  truthfully, provided that all other consumers report their types truthfully. Therefore, from the viewpoint of  the $i$th consumer at time $t$ BIC constraint can be expressed as follows: 
\begin{equation}
\begin{split}
&\hspace{-1em}\mathbb{E}_{H_t^{-i}}\Big[ \; \tilde{\theta}^i_t \:  \sum\limits_{j \le \tilde{b}^i_t} q^{i,j}_{t}(H_t^{-i}, (\tilde{\theta}^i_t, \tilde{b}^i_t))  - p_{t}^{i}(H_{t}^{-i}, (\tilde{\theta}^i_t, \tilde{b}^i_t)) \;  | \; N_t = n_t \; \Big] \: \\
&\hspace{-1em}\ge 
\mathbb{E}_{H_{t}^{-i}}\Big[ \; \tilde{\theta}^i_t \:  \sum\limits_{j \le \tilde{b}^i_t} q^{i,j}_{t}(H_{t}^{-i}, (r^i_t, c^i_t)) - p_{t}^{i}(H_{t}^{-i}, (r^i_t, c^i_t)) \;   | \; N_t = n_t \; \Big], \\
&\forall \tilde{\theta}^i_t , r^i_t \in \Theta  \; , c^i_t \le \tilde{b}^i_t, \; c^i_t, \tilde{b}^i_t \in \{1, 2, \ldots, k\} \; , \forall n_t, \forall t.
\end{split}
\raisetag{1\baselineskip}
\label{BIC}
\end{equation}

Individual Rationality (IR) constraint ensures that the consumer's expected utility at the truthful reporting equilibrium is non-negative. Using \eqref{truthEU}, from the viewpoint of  $i$th consumer at time $t$ IR constraint  can be described as follows:
\begin{equation}
\begin{split}
&\mathbb{E}_{H_{t}^{-i}}\Big[ \; \tilde{\theta}^i_t \:  \sum\limits_{j \le \tilde{b}^i_t} q^{i,j}_{t}(H_{t}^{-i}, (\tilde{\theta}^i_t, \tilde{b}^i_t))  \\
&- p_{t}^{i}(H_{t}^{-i}, (\tilde{\theta}^i_t, \tilde{b}^i_t)) \;   | \; N_t = n_t \; \Big] \: \ge 0 \; , \; \forall n_t, \; \forall t. 
\end{split}
\raisetag{1\baselineskip}
\label{IR}
\end{equation}

\subsection{Expected Revenue Maximization} \label{sec:revMax}
Consider a BIC and IR mechanism $(q_{1:T}, p_{1:T})$. When all consumers adopt the truthful strategy, the history at time $t$ is
\begin{align}
H_{t} := \Big\{ &\{(\theta_1^i, b_1^i)\}_{i=1}^{N_1}, \ldots, \{(\theta_t^i, b_t^i)\}_{i=1}^{N_t}, 
Y^{1:k}_{1:t},  X^{1:k}_{1:t} \Big\}, \label{eq:Ht}
\end{align}
and the expected total revenue is $\mathbb{E}\Big\{ \sum\limits_{t=1}^T  \sum\limits_{i =1}^{N_t} p^{i}_{t}(H_{t}) \Big\}$. 



The mechanism design problem  can now be formulated as
\begin{equation} \label{auc-opt}
\begin{split}
\max\limits_{(q_{1:T}, p_{1:T})} \; \; \; \mathbb{E}\Big\{ \sum\limits_{t=1}^T  \sum\limits_{i =1}^{N_t} p^{i}_{t}(H_{t}) \Big\} \;  , 
\; \text{subject to} \; \; \text{\eqref{BIC}  , \eqref{IR}}.
\end{split}
\end{equation}

\section{Characterization of BIC and IR Mechanisms}\label{sec:MechChar}
In this section we provide a  characterization of BIC and IR mechanisms that will be useful for solving the problem in \eqref{auc-opt}.
\subsection{Interim Allocation and Payment}\label{sec:interim}
Suppose $n_t$ consumers arrive at time $t$ and let $(\tilde{\theta}^i_t, \tilde{b}^i_t)$ be the true type of the $i$th consumer arriving at time $t$.  Assuming that all other consumers report their types truthfully,    this consumer's expected allocation and payment under the mechanism $(q_{1:T}, p_{1:T})$ if it reports the pair  $(r^i_t, c^i_t)$  are given as 
\begin{align} 
Q^i_t( r^i_t , c^i_t , n_t) &\coloneqq \mathbb{E}_{H^{-i}_{t}}\Big[ \sum\limits_{j \le c^i_t} q_{t}^{i,j}(H^{-i}_{t}, (r^i_t, c^i_t)) \;   | \; N_t = n_t \Big] \label{eq:interim} \\
P^i_t( r^i_t, c^i_t, n_t) &\coloneqq \mathbb{E}_{H^{-i}_{t}}\Big[ \; p^{i}_{t}(H^{-i}_{t}, (r^i_t, c^i_t)) \;   | \;  N_t= n_t  \Big]
\label{eq:P}
\end{align}
In the following lemmas,  we provide an operational characterization of the BIC and IR mechanisms in terms of the interim quantities defined in (\ref{eq:interim})-(\ref{eq:P}). 
\begin{lemma}\label{lem:BICIR}
A mechanism $(q_{1:T},p_{1:T})$ satisfies the BIC and IR constraints if given $N_t = n_t$,  the following conditions hold true for all $i \in \{1, \ldots, n_t\}, \forall t,$
\begin{enumerate}[(i)]
\item $Q^i_{t}( r, c, n_t)$ is non-decreasing in $r$ for all $c \in \{1,\ldots,k\}.$
\item $Q^i_{t}( r, c, n_t)$ is non-decreasing in $c$ for all $r \in \Theta.$
\item $P^i_{t}(\theta^{\text{min}}, c, n_t) = 0\;$, for all $\; c \in \{1, 2, \ldots, k\}$.
\item The interim payment  in (\ref{eq:P}) takes the following form for all  $r \in \Theta, c \in \{1, 2, \ldots, k \}$: \vspace{-1.3em}
\begin{align}\label{eq:interimP}
P^i_{t}( r, c, n_t) &= r \; Q^i_{t}( r, c, n_t) - \int\limits_{\theta^{\text{min}}}^{r} Q^i_{t}(s, c, n_t) \; ds \\
&- \theta^{\text{min}} \; Q^i_{t}(\theta^{\text{min}}, c, n_t). \notag
\end{align}
\item $\theta^{\text{min}} \; Q^i_{t}(\theta^{\text{min}}, c, n_t) \ge 0 \;, \forall c$.
\end{enumerate}
\begin{proof}
The proof closely follows the standard arguments in  \cite[Section III]{navabi2018optimal}.
\end{proof}
\end{lemma}


\begin{lemma}\label{lem:BICthetaIR}
Any BIC and IR  mechanism $(q_{1:T},p_{1:T})$ satisfies 
\begin{align}\label{eq:interimPineq}
&P^i_{t}( r, c, n_t) \le r \; Q^i_{t}( r, c, n_t) - \int\limits_{\theta^{\text{min}}}^{r} Q^i_{t}(s, c, n_t) \; ds, \\
&\text{for all} \;\; t, n_t,  i \in \{1, \ldots, n_t\},  r \in \Theta, c \in \{1, 2, \ldots, k \}. \notag
\end{align}
\begin{proof}
The proof closely follows the standard arguments in  \cite[Section III]{navabi2018optimal}.
\end{proof}
\end{lemma}

\section{Revenue Maximizing Mechanism}\label{sec:rev}
In this section we  characterize  the expected-revenue maximizing mechanism. 
Let us define
\begin{equation}\label{eq:virVal}
w_t(\tilde{\theta}, \tilde{b}) \coloneqq \Big( \tilde{\theta} - \frac{1 - \Pi_t(\tilde{\theta} \; | \; \tilde{b})}{\pi_t(\tilde{\theta} \; | \; \tilde{b})}  \Big),
\end{equation}
where $\Pi_t(\cdot \; | \; \tilde{b})$ is the  cumulative distribution function (CDF) corresponding to the conditional  probability density function (pdf) $\pi_t(\cdot \; | \; \tilde{b})$. In economics terminology, $w_t(\tilde{\theta}, \tilde{b})$ is referred to as the \textit{virtual valuation} \cite[Chapter 3]{borgers2015introduction} of a consumer with type $(\tilde{\theta}, \tilde{b})$ that arrives at time $t$.

We make the following assumptions to simplify the solution to the optimal mechanism design problem in \eqref{auc-opt}.
\begin{assum}\label{assum:mhrc}
We assume that
\begin{enumerate}[(i)]
\item The conditional probability density  functions $\pi_t(\cdot | c), t = 1, \ldots, T, c = 1, \ldots, k$ satisfy the generalized monotone hazard rate condition \cite[Section 2]{pai2013optimal}, \cite[Section IV]{navabi2018optimal}. That is, for all $t$, we assume $\frac{\pi_t(x | c)}{1-\Pi_t(x | c)}$ is non-decreasing in $x$ and $c$. Moreover, we assume that for all $t$ if $x \ge x'$ and $c > c'$, then $\frac{\pi_t(x | c)}{1-\Pi_t(x | c)} > \frac{\pi_t(x' | c')}{1-\Pi_t(x' | c')}$. \\
\item $w_t(\theta^{\text{min}}, j) < 0$ for all $\; j , t$.
\end{enumerate}
\end{assum}
In the following lemma we provide a characterization of the expected-revenue maximizing mechanism.
\begin{lemma}\label{lem:revEqBern}
Suppose $(q^*_{1:T}, p^*_{1:T})$ is a BIC and IR mechanism for which  the following conditions are true:
\begin{enumerate}[(i)]
\item $(q^*_{1:T})$ is the solution to the following functional  optimization problem 
\begin{equation}\label{revenue}
\max\limits_{q_{1:T}} \; \mathbb{E}\Big[ \;   \sum\limits_{t=1}^T   \sum\limits_{i=1}^{N_t} w_t(\theta_t^i, b_t^i) \; \Big(  \sum\limits_{j \le b_t^i} q_{t}^{i, j}(H_{t}) \Big)  \; \Big],
\end{equation}
where $H_t$ is the history under truthful reporting.
\item Given the history $h_t$ and assuming that $n_t$ consumers arrive at time $t$, the payment charged to  the $i$th consumer arriving at time $t$ with the true type $(\tilde{\theta}_t^i, \tilde{b}_t^i)$ is given as: 
\begin{align}\label{eq:expostp}
&\hspace{-2em}p^{*i}_t(h^{-i}_t, (\tilde{\theta}_t^i, \tilde{b}_t^i)) = \tilde{\theta}_t^i \sum\limits_{j \le \tilde{b}_t^i} q_t^{*i,j}(h_t^{-i}, (\tilde{\theta}_t^i, \tilde{b}_t^i)) \notag \\
&\hspace{-2em}- \int\limits_{\theta^{\text{min}}}^{\tilde{\theta}_t^i} \Big( \sum\limits_{j \le \tilde{b}_t^i} q_t^{*i,j}(h_t^{-i}, (\alpha, \tilde{b}_t^i)) \Big) \; d\alpha , \forall i \in \{1, \ldots, n_t\}, \forall n_t, \forall t, 
\raisetag{3\baselineskip}
\end{align}
where, $h_t^{-i} = h_t \setminus \{(\tilde{\theta}_t^i, \tilde{b}_t^i)\}$.
\end{enumerate}
 Then $(q^*_{1:T}, p^*_{1:T})$ gives the highest expected revenue in the class of BIC and IR mechanisms. 
\begin{proof}
See Appendix \ref{sec:revEqBern-pf}.
\end{proof}
\end{lemma} 
The results of Lemma \ref{lem:revEqBern}  imply that in order  for a BIC and IR mechanism to maximize the expected revenue, its  allocation rules must  solve the functional optimization problem in \eqref{revenue}. This problem can be viewed as a stochastic control problem. In the following section, we describe this stochastic control problem and formulate a dynamic program to find the optimal allocation rules.

\section{Solution to the Stochastic Control Problem}\label{sec:ctrlSol}
The optimization problem in \eqref{revenue} is a finite horizon stochastic control problem with the history at time $t$ (with truthful reporting) as the state and the allocation matrix as the action at time $t$. The allocation functions $q_{1:T}$ are the control strategy and the optimization in \eqref{revenue} is to find the control strategy with the highest expected reward. This stochastic control perspective provides a dynamic program for the optimization in \eqref{revenue}.  We then leverage the  structure of the consumers' flexibility model  to simplify the dynamic program.
\subsection{Dynamic Program}\label{sec:DP}
For a truthful history $h_t$ at time $t$, let $R_t(h_t)$ denote the maximum expected reward from $t$ to $T$ for the stochastic control problem in \eqref{revenue}. $R_t(h_t)$ is a value function and obeys the standard dynamic programming recursions given below:
\begin{align}
\text{If} \;\; h_t^R = \emptyset:& \;\;\; R_t(h_t) \coloneqq \mathbb{E}\Big[R_{t+1}(H_{t+1}) \; | \; h_t \Big]  \label{eq:DPqtempty} \\
 \text{If} \;\; h_t^R \neq \emptyset:& \; \notag \\ 
R_t(h_t) &\coloneqq \max\limits_{\bold{A} \in \mathcal{S}(h^R_t, y_t^{1:k})} \Big\{ \;  \sum\limits_{i =1}^{|h^R_t|} w_t(  \tilde{\theta}_t^i, \tilde{b}_t^i) \sum\limits_{j=1}^k \bold{A}(i,j) \label{eq:DPqt} \\
&+ \mathbb{E}\Big[R_{t+1}(H_{t+1}) \;|\; h_t, \bold{A}_t = \bold{A} \Big] \Big\}, \notag
\raisetag{1\baselineskip}
\end{align}
where $R_{T+1}(\cdot) = 0$.

In the above dynamic program, the information state at time $t$ is $h_t$ (since the value functions have $h_t$ as the argument). It can be shown that the only relevant part of the history are the reports and the state of supply at current time. In the following lemma, we use this idea to simplify the information state and the dynamic program. 
\begin{lemma}\label{lem:infoState}
Let $s_t =(h^R_t, y^{1:k}_t)$. There exist functions $V_1(\cdot), \ldots, V_T(\cdot)$ such that at each time $t$:
\begin{align}\label{eq:Vfuncs}
&V_t(s_t) = R_t(s_t, x_t^{1:k}, h_{t-1}) \; , \notag \\
&\text{for all} \; \{ s_t, x_t^{1:k}, h_{t-1} \} \in \mathcal{H}_t.
\end{align}
Further, these functions obey the following dynamic program:
\begin{align}
\text{If} \;\; h_t^R = \emptyset:& \;\;\; V_t(s_t) = \mathbb{E}\Big[V_{t+1}(S_{t+1}) \;|\; s_t \Big]  \label{eq:DPVqtempty} \\
 \text{If} \;\; h_t^R \neq \emptyset:& \; \notag \\ 
V_t(s_t) &= \max\limits_{\bold{A} \in \mathcal{S}(h^R_t, y_t^{1:k})} \Big\{ \;  \sum\limits_{i =1}^{|h^R_t|} w_t(  \tilde{\theta}_t^i, \tilde{b}_t^i) \sum\limits_{j=1}^k \bold{A}(i,j) \label{eq:DPVqt} \\
&+ \mathbb{E}\Big[V_{t+1}(S_{t+1}) \;|\; s_t, \bold{A}_t = \bold{A} \Big] \Big\}, \notag
\raisetag{1\baselineskip}
\end{align}
where $V_{T+1}(\cdot) = 0$.
\begin{proof}
See Appendix \ref{sec:infoStatePf}.
\end{proof}
\end{lemma}


Based on the results of  Lemma \ref{lem:infoState}, the optimal allocation functions can be described in terms of the solution to the  dynamic program in  (\ref{eq:DPVqtempty})-(\ref{eq:DPVqt}) with the simplified information state $s_t  = (h^R_t, y_t^{1:k})$ at  time $t$. 
 
In the  dynamic program in (\ref{eq:DPVqt}) at each time $t$, the optimization variables comprise all the entries of the $|h^R_t|\times k$  allocation matrix $\bold{A}_t$. In sequel we simplify the dynamic program formulation in terms of  alternate  variables that need to be optimized at each time.
\subsection{Alternative Optimization Variables}\label{sec:varuv}
Consider  the information state $(h^R_t, y_t^{1:k})$ with $h^R_t \neq \emptyset$ and an allocation matrix $\mathbf{A} \in \mathcal{S}(h^R_t, y_t^{1:k})$ at time $t$. Let $u^j_t$  denote the number of consumers with flexibility level $j$ that get a good at time $t$ under  $\mathbf{A}$. That is, $u^j_t := \sum\limits_{\substack{i=1 \\ i: \tilde{b}^i_t = j}}^{|h^R_t|} \sum\limits_{l \le j} \mathbf{A}(i,l)$, where $\mathbf{A}(i,l)$ is the entry in the $i$th row and $l$th column of the matrix $\mathbf{A}$.   Let $\mathcal{U}(h^R_t, y_t^{1:k})$ denote the set of admissible values of $u^{1:k}_t$ given the information state $(h^R_t, y_t^{1:k})$. Thus  for every vector $u^{1:k} \in \mathcal{U}(h^R_t, y_t^{1:k})$, there exists a matrix $\hat{\bold{A}} \in \mathcal{S}(h^R_t, y_t^{1:k})$ such that $\sum\limits_{\substack{i=1 \\ i: \tilde{b}^i_t = j}}^{|h^R_t|} \sum\limits_{l \le j}\hat{\bold{A}}(i,l) = u^j, \forall j$. If $h^R_t = \emptyset$, no consumer is present to be allocated and thus we define $\mathcal{U}(h^R_t, y_t^{1:k}) := \{\bold{0}_{1\times k}\}$. In the following lemma, we provide a more operational characterization  of $\mathcal{U}(h^R_t, y_t^{1:k})$.
\begin{lemma}\label{lem:Uset}
Given the information state $(h^R_t, y_t^{1:k})$ at time $t$, 
\begin{align}\label{eq:Uset}
\mathcal{U}(h^R_t, y_t^{1:k}) = \Big\{ u^{1:k} \in \mathbb{Z}^k_{\ge 0} :  \sum\limits_{l=1}^j u^l \le \sum\limits_{l=1}^j y_t^l \; , \;  u^j \le n_t^j, \forall j  \Big\},
\end{align}
where  $n_t^j$ denotes the number of consumers with flexibility level $j$ that arrive at time $t$.
\begin{proof}
See Appendix \ref{sec:UsetPf}.
\end{proof}
\end{lemma}
Consider the information state $(h^R_t, y_t^{1:k})$ with $h^R_t \neq \emptyset$ and an allocation matrix $\mathbf{A} \in \mathcal{S}(h^R_t, y_t^{1:k})$ at time $t$. Let $v^j_t$  denote the  number of goods of variety $j$ allocated at time $t$ under $\mathbf{A}$, i.e.,  $v^j_t \coloneqq \sum\limits_{i=1}^{|h^R_t|} \mathbf{A}(i,j)$.  Given some  vector $u_t^{1:k} \in \mathcal{U}(h^R_t, y_t^{1:k})$, let $\mathcal{V}(u^{1:k}_t, y_t^{1:k})$ denote the set of all  values of  $v^{1:k}_t$ that can fulfill the demand represented by $u^{1:k}_t$ under the supply $y_t^{1:k}$. More precisely,  for every vector $v^{1:k} \in \mathcal{V}(u^{1:k}_t, y_t^{1:k})$, there exists a matrix $\hat{\bold{A}} \in \mathcal{S}(h^R_t, y_t^{1:k})$ such that for each $j$, $\hat{\bold{A}}$ serves $u^j_t$ consumers of flexibility level $j$ (i.e., $\sum\limits_{\substack{i = 1 \\ i:\tilde{b}_t^i = j}}^{|h_t^R|} \sum\limits_{l\le j} \hat{\bold{A}}(i,l) = u_t^j$)  and allocates $v^j$ goods of variety $j$ (i.e., $\sum\limits_{i=1}^{|h_t^R|} \hat{\bold{A}}(i,j) = v^j$). If $h^R_t = \emptyset$, no consumer is present to be allocated and thus $\mathcal{V}(u_t^{1:k}, y_t^{1:k}) = \mathcal{V}(\bold{0}_{1\times k}, y_t^{1:k}) = \{\bold{0}_{1\times k}\}$. In the following lemma, we provide a more operational characterization of $\mathcal{V}(u^{1:k}_t, y_t^{1:k})$.
\begin{lemma}\label{lem:Vset}
Given the information state $(h^R_t, y_t^{1:k})$ and the vector $u_t^{1:k} \in \mathcal{U}(h^R_t, y_t^{1:k})$ at time $t$, 
\begin{align}\label{eq:Vset}
&\mathcal{V}(u^{1:k}_t, y_t^{1:k}) = \Big\{ v^{1:k} \in \mathbb{Z}_{\ge 0}^k : \\
&v^j \le y^j_t \; , j = 1, \ldots, k, \notag \\
&\sum\limits_{l=1}^j u_t^l \le \sum\limits_{l=1}^j v^l \; , \; j = 1, \ldots, k-1 \; , \; \sum\limits_{l=1}^k u_t^l = \sum\limits_{l=1}^k v^l \; \Big\}. \notag
\end{align}
\begin{proof}
The proof is similar to the proof of Lemma \ref{lem:Uset} and is therefore omitted.
\end{proof}
\end{lemma} 
In the following lemma we show that  the two $k$-dimensional vectors $u_t^{1:k}$ and $v_t^{1:k}$ constructed above,  can be treated as  the   optimization variables  in the dynamic program in (\ref{eq:DPVqt}). 
\begin{lemma}\label{lem:DPuv}
The value function in  (\ref{eq:DPVqt}) can be  equivalently expressed below:
\begin{align}
&V_t(h^R_t, y_t^{1:k}) = \max\limits_{u^{1:k} \in \mathcal{U}(h^R_t, y_t^{1:k})} \Big\{ \;  \sum\limits_{j=1}^{k} \sum\limits_{i=1}^{u^j} w_t^{i,j} \notag \\
&+ \max\limits_{v^{1:k} \in \mathcal{V}(u^{1:k}, y_t^{1:k})} \mathbb{E}\Big[V_{t+1}\Big(H^R_{t+1}, \{y^j_t - v^j + X^j_{t+1}\}_{j=1}^k \Big)  \Big] \Big\},   \label{eq:DPuvt}
\raisetag{3\baselineskip}
\end{align} 
where $V_{T+1}(\cdot)=0$. In  (\ref{eq:DPuvt}) $w_t^{i,j}$ denotes the $i$th largest element in $\mathcal{W}_t^j$ defined below:
\begin{align}\label{eq:Wjt}
\mathcal{W}_t^j \coloneqq \{ \; w_t(\tilde{\theta}, \tilde{b}) : (\tilde{\theta}, \tilde{b}) \in h^R_t \; , \tilde{b} = j \; \},
\end{align}
that is, $\mathcal{W}_t^j$ denotes the set of virtual valuations (see (\ref{eq:virVal})) of all the consumers with flexibility level $j$ at time $t$.
\begin{proof}
See Appendix \ref{sec:DPuvPf}.
\end{proof}
\end{lemma}
In the following lemma, we  establish a  monotonicity property of the value functions in (\ref{eq:DPuvt}). In Lemma \ref{lem:v*recipe}    we  leverage this property to construct an optimal solution to the inner maximization over $v^{1:k}$ vector  in (\ref{eq:DPuvt}).
\begin{lemma}\label{lem:Rmonotone}
Consider  two supply profiles $y_t^{1:k}$ and $z_t^{1:k}$ such that:
\begin{equation}\label{yz}
\begin{split}
y^i_t &= z^i_t + 1 \\
y^j_t &= z^j_t -1 \\
y_t^l &= z_t^l \; , \; \; \text{for all} \; \; l \neq i , j \; ,
\end{split}
\end{equation}
where $i<j$. 
Then, the value functions $V_t(\cdot)$ defined in Lemma \ref{lem:infoState} satisfy the following property:
\begin{align}
&V_t(h^R_t, y_t^{1:k})  \ge V_t(h^R_t, z_t^{1:k}) \; , \; \text{for all $h^R_t$ and $t$}. \label{eq:def1Rmon}
\end{align} 
\begin{proof}
See Appendix \ref{sec:yzPf}.
\end{proof}
\end{lemma}
A more intuitive interpretation  of the property established in Lemma \ref{lem:Rmonotone} is that  a good of  variety $j$ contributes more to the generation of revenue in comparison with a good of any of the varieties $j+1, j+2, \ldots, k$. Consequently, allocating a  good of  variety $j$ is at least as costly as allocating a good of any of the varieties $j+1, j+2, \ldots, k$.

\begin{remark}\label{rem:generalMonPropOfV}
The value functions $V_t(\cdot)$ defined in Lemma \ref{lem:infoState} satisfy a more general version of the monotonicity property shown in Lemma \ref{lem:Rmonotone}. Consider any two supply profiles $y_t^{1:k}$ and $z_t^{1:k}$ such that: $\sum_{l=1}^j y^l_t \ge \sum_{l=1}^j z^l_t, j = 1, \ldots, k$. Then, it can be shown that value functions $V_t(\cdot)$ still  satisfy (\ref{eq:def1Rmon}).
\end{remark}

The property  in (\ref{eq:def1Rmon}) can be leveraged to provide an explicit solution for the inner maximization over $v^{1:k}$ in (\ref{eq:DPuvt}). This is shown in Lemma \ref{lem:v*recipe}.
\begin{lemma}\label{lem:v*recipe}
At each time $t$, given $h^R_t, y_t^{1:k}$ and $u^{1:k} \in \mathcal{U}(h^R_t, y_t^{1:k})$, \underline{recursively} define
\begin{align}
v^{*k} &:= \min(u^k, y_t^k) \label{eq:v*k}\\
v^{*j} &:= \min\Big(y_t^j,  u^j + (\sum\limits_{l=j+1}^k u^l - \sum\limits_{l=j+1}^k v^{*l} ) \Big), j = k-1, \ldots, 1. \notag
\end{align}
Then, 
\begin{align}\label{eq:v*}
&v^{*1:k} \in \argmax\limits_{v^{1:k} \in \mathcal{V}(u^{1:k}, y_t^{1:k})} \mathbb{E}\Big[ V_{t+1}\Big(H^R_{t+1}, \{y^j_t - v^j + X^j_{t+1}\}_{j=1}^k \Big) \Big]. 
\end{align}
\begin{proof}
See Appendix \ref{sec:v*recipePf}.
\end{proof}
\end{lemma}
Based on the results of  Lemma \ref{lem:v*recipe}, the value functions in (\ref{eq:DPuvt}) can be simplified as  expressed below:
\begin{align}
V_t(h^R_t, y_t^{1:k}) &= \max\limits_{u^{1:k} \in \mathcal{U}(h^R_t, y_t^{1:k})} \Big\{ \;  \sum\limits_{j=1}^{k} \sum\limits_{i=1}^{u^j} w_t^{i,j} \notag \\
&+ \mathbb{E}\Big[V_{t+1}\Big(H^R_{t+1}, \{y^j_t - v^{*j} + X^j_{t+1} \}_{j=1}^k \Big)  \Big] \Big\}, \label{eq:DPuv*t} 
\raisetag{3\baselineskip}
\end{align} 
where  $v^{*1:k}$ is obtained corresponding to each  $u^{1:k} \in \mathcal{U}(h^R_t, y_t^{1:k})$ as described in (\ref{eq:v*k}). 
\section{The Optimal Mechanism} \label{sec:OPTmech}
In Lemma \ref{lem:revEqBern}, we established that the expected-revenue maximizing allocation rules of a BIC and IR mechanism indeed coincide with the optimal control strategy for the  stochastic control problem in \eqref{revenue}. Based on this insight,  Section \ref{sec:ctrlSol} was devoted to development of a  characterization of the optimal control strategy for the problem in  \eqref{revenue} in terms of the solution to a  dynamic program (see (\ref{eq:DPqtempty})-(\ref{eq:DPqt})). We  leveraged the structure of the flexibility model to simplify the  formulated dynamic program (see lemmas \ref{lem:DPuv}-\ref{lem:v*recipe}).   In the following theorem, we use the results of lemmas \ref{lem:revEqBern}-\ref{lem:v*recipe} as well as the characterization of  BIC and IR mechanisms provided in Lemma \ref{lem:BICIR} to specify the allocation and payment rules of the optimal mechanism.
\begin{theorem}\label{thm:opt-mech}
Consider the information state $(h^{-i,R}_t, (r,j), y_t^{1:k})$, where  consumer $i$ reports $(r,j)$ as its type  and  $h^{-i,R}_t$ denotes the set of reports from all  consumers other than $i$.
Let $u_t^{*1:k}$ and $v_t^{*1:k}$  denote the optimal vectors that result from solving the dynamic program  in (\ref{eq:DPuv*t}). Consider the mechanism $(q^*_{1:T}, p^*_{1:T})$ described below:

\begin{itemize}
\item \textbf{Allocations:} Let  $q^*_t(h^{-i,R}_t, (r,j), y_t^{1:k}) \in \mathcal{S}(h^{-i,R}_t, (r,j), y_t^{1:k})$ denote the allocation matrix  constructed according to the  allocation procedure described below:
\bromalist
\item Index the  goods under the profile $v_t^{*1:k}$   in a non-decreasing flexibility order, i.e., the $v_t^{*1}$ goods of variety 1   are indexed as $1, \ldots, v_t^{*1}$ and, the $v_t^{*2}$ goods of variety 2   are indexed as $v_t^{*1}+1, \ldots, v_t^{*1}+v_t^{*2}$, and  so on.
\item Sort consumers of flexibility level  1 in non-increasing order of virtual valuations. Top $u_t^{*1}$ consumers of flexibility level  1 get  the first $u_t^{*1}$ goods as arranged in \textit{(i)}. Ties are resolved randomly.
\item Sort consumers of flexibility level 2 in non-increasing order of virtual valuations. Top $u_t^{*2}$ consumers of flexibility level  2  get  the next $u_t^{*2}$ goods. Ties are resolved randomly.
\item Allocations to the top $u_t^{*j}$ consumers with flexibility levels $j = 3, \ldots, k$ are carried out in the same fashion as above. 
\item The rest of the consumers do \underline{not} get an allocation. 
\end{list}
\item \textbf{Payments:} Suppose  $n_t$ consumers arrive at time $t$. The payment function $p_t^{*}(h^{-i,R}_t, (r,j), y_t^{1:k}) \in \mathbb{R}^{n_t}$ is  defined below for $i =1, \ldots, n_t$:
\begin{align}
&p_t^{*i}(h^{-i,R}_t, (r,j), y_t^{1:k})  \notag \\
&= \left\{ \begin{array}{ll}
\bar{\theta}^{i,j}_t &\mbox{if consumer $i$ gets a good}    \\ 
0 &\mbox{otherwise}
\end{array}\right.,  \label{eq:p*} 
\end{align}
where   $\bar{\theta}^{i,j}_t$ is defined as 
\begin{align}
&\bar{\theta}^{i,j}_t \coloneqq \sup\Big\{ x \in [\theta^{\text{min}}, \theta^{\text{max}}] : \notag \\
&\sum\limits_{l \le j} q^{*i,l}_t(h^{-i,R}_t , (x,j), y^{1:k}_t) = 0 \; , \;  w_t(x,j) \ge 0 \Big\}. \label{eq:thetabarijt}
\raisetag{1.5\baselineskip}
\end{align}
\end{itemize}
Under Assumptions \ref{assum:typeAs}-\ref{assum:mhrc}, $(q^*_{1:T}, p^*_{1:T})$ is an expected-revenue maximizing, BIC and IR mechanism. 
\begin{proof}
See Appendix \ref{sec:opt-mech-pf}.
\end{proof}
\end{theorem}

\begin{remark}\label{rem:dynSolNumeric}
The  vectors $u_t^{*1:k}$ and $v_t^{*1:k}$ that characterize the optimal allocation matrix $q^*_t(\cdot)$  in Theorem \ref{thm:opt-mech} as well as the    quantities $\bar{\theta}^{i,j}_t$ defined in (\ref{eq:thetabarijt})   can be found through discretizing the set $[\theta^{\text{min}}, \theta^{\text{max}}]$ with sufficient numerical precision and applying the methods developed for solving Markov decision processes with discrete state-action spaces, such as policy iteration, value iteration and linear programming  \cite{puterman2014markov}.
\end{remark}

\subsection{Example}\label{ex:DPsol}
Consider a simple setup with $T=2$ and $k=2$,  where the consumer arrival process $\lambda_t(\cdot)$ follows a Bernoulli distribution, that is, at each time step a consumer may enter the market with probability $p$.  For a consumer entering the market at time $t=1,2$, its flexibility level is equally likely to be 1 or 2, i.e., $g_t(j)=\frac{1}{2}$ for $j \in \{1,2\}$ and, conditioned on its flexibility level being $j$, its valuation has truncated exponential distribution over the interval $[0,1]$, i.e., $\pi_t(x|j) = \frac{\alpha_j \exp(-\alpha_j x)}{1 - \exp(-\alpha_j)}, x \in [0,1]$, where $\alpha_2 > \alpha_1 > 0$. It is straightforward to verify that $\pi_t(\cdot|j)$ satisfies Assumption \ref{assum:mhrc}. The virtual valuation function (see (\ref{eq:virVal})) associated with $\pi_t(\cdot|j)$ is of the following form:
\begin{align*}
w_t(x,j) = x - \frac{1}{\alpha_j}\Big(1 - \exp(\alpha_j(x-1))\Big) \; , \; t=1,2.
\end{align*}
Suppose the supply profile at time $t=1$ is $(y^1_1,y^2_1) = (1,1)$ and   no more goods of either variety becomes available at time $t=2$. Suppose a consumer with type $(\tilde{\theta}, 2)$ arrives at time $t=1$. Under the optimal mechanism characterized in Theorem \ref{thm:opt-mech}, this consumer gets a good of variety 2 if $w_1(\tilde{\theta}, 2) > \mathbb{E}[V_2(H^R_2, (1,1)) - V_2(H^R_2, (1,0))] =: \rho^2_1$ (see (\ref{eq:DPuv*t})). It is straightforward to verify that in this example
\begin{align*}
&\mathbb{E}[V_2(H^R_2, (1,1))] = \mathbb{E}[V_2(H^R_2, (1,0))] \\
&= \frac{p}{2} \Big(\mathbb{E}_{\theta|b=1}\Big[\max\{w_2(\theta, 1),0\}\Big]+\mathbb{E}_{\theta|b=2}\Big[\max\{w_2(\theta, 2),0\}\Big]\Big),
\end{align*}
from which it follows that $\rho^2_1 = 0$.

If instead a consumer  with type  $(\tilde{\theta}, 1)$ arrives at time $t=1$, it gets a good of variety 1 if $w_1(\tilde{\theta}, 1) > \mathbb{E}[V_2(H^R_2, (1,1)) - V_2(H^R_2, (0,1))] =: \rho^1_1$. We observe that in this example
\begin{align*}
&\mathbb{E}[V_2(H^R_2, (0,1))] = \frac{p}{2} \mathbb{E}_{\theta|b=2}\Big[\max\{w_2(\theta, 2),0\}\Big]
\end{align*}
and thus,
\begin{align}
\rho^1_1 &= \frac{p}{2} \mathbb{E}_{\theta|b=1}\Big[\max\{w_2(\theta, 1),0\}\Big] \notag \\
&= \frac{p}{2} \theta^{\text{res}}_{1,2} \frac{\exp(-\alpha_1 \theta^{\text{res}}_{1,2}) - \exp(-\alpha_1)}{1 - \exp(-\alpha_1)},
\label{eq:rho11}
\end{align}
where $\theta^{\text{res}}_{j,t}$ is as defined below
\begin{align}\label{eq:theta-res-j}
\theta^{\text{res}}_{j,t} := \max\Big\{ x \in [0,1] : w_t(x,j) = 0 \Big\}.
\end{align}
For instance if $\alpha_1 = 2$ and $p=0.5$ we obtain $\theta^{\text{res}}_{1,2} \approx 0.36$ and thus $\rho^1_1 \approx 0.037$. 

In general, for the setup described above, it is easy to check that  $\rho^1_1 \ge \rho^2_1$, which implies that a consumer with  flexibility level 1 needs to have higher valuation to get an allocation at $t=1$. If the consumer with type $(\tilde{\theta}, j)$ at time $t=1$ gets a good, it pays $\bar{\theta}^j_1 = w_1^{-1}(\rho^j_1  ; j)$ (see (\ref{eq:thetabarijt})), where $w_1^{-1}(\cdot ; j)$ denotes inverse of $w_1(\cdot , j)$.  Notice that  Assumption \ref{assum:mhrc} combined with  $\rho^1_1 \ge \rho^2_1$ implies that $\bar{\theta}^1_1 \ge \bar{\theta}^2_1$, i.e., a consumer with  flexibility level 1 is charged a higher price upon allocation of a desired good.    For instance, consider $\alpha_2 = 3, \alpha_1 = 2$ and $p=0.5$. For this numerical  setup we obtain  $\bar{\theta}^1_1 = w_1^{-1}(\rho^1_1  ; 1) \approx w_1^{-1}(0.037  ; 1) \approx 0.39$ and $\bar{\theta}^2_1 = w_1^{-1}(\rho^2_1  ; 2) = w_1^{-1}(0  ; 2)  \approx 0.29$.

At time $t=2$ which is the terminal time step, if a consumer with type $(\tilde{\theta}, j)$ arrives and a desired good is available, it gets an allocation if $w_2(\tilde{\theta}, j) > 0$ and is charged  the reserve price  $\theta^{\text{res}}_{j,t}$ at $t=2$ as defined in (\ref{eq:theta-res-j}).  
Notice that from Assumption \ref{assum:mhrc} it follows that $\theta^{\text{res}}_{1,2} > \theta^{\text{res}}_{2,2}$, meaning that at time $t=2$ also, a consumer with  flexibility level  1 needs to have a higher valuation to get an allocation and  is charged a higher price upon allocation of a desired good. For instance for the case $\alpha_2 = 3, \alpha_1 = 2$ and $p=0.5$ we see that $\theta^{\text{res}}_{1,2} \approx 0.36  > \theta^{\text{res}}_{2,2} \approx 0.29$.

Therefore we observe that in the above setup under the optimal mechanism, at each time the payment charged to the more flexible consumers is less than the payment charged to the less flexible consumers. Moreover, it is straightforward to verify that $\bar{\theta}^j_t$ is non-increasing  in $t$, that is, the payment charged to the consumers with flexibility level $j$ decreases over time across  all $j$ (e.g., for the case $\alpha_1 = 2 , p=0.5$ we see that $\bar{\theta}^1_1 \approx 0.39 > \bar{\theta}^1_2 = \theta^{\text{res}}_{1,2} \approx 0.36$). Intuitively, this is expected   because unlike  $t=1$, the consumer that arrives at $t=2$, faces no competition from future  consumers. On the other hand, the virtual valuation of a consumer  that arrives at $t=1$ must outweigh the expected revenue that can  be produced by saving the good for   a consumer that may arrive at $t=2$. As a result, a consumer that arrives  at  $t=2$  is expected to be charged less (only the reserve price associated with its flexibility level) than a consumer of the same flexibility level that arrives at $t=1$. 

\section{Conclusion}\label{sec:conclusion}
We studied the problem of designing a dynamic expected-revenue maximizing, BIC and IR mechanism for allocation of multiple goods of $k$ varieties to flexible consumers over $T$ time steps. In our model, a random number of goods of each variety may become available to the seller at each time and a random number of consumers may enter the market at each time. We considered impatient consumers that need to get one good of one of their desired varieties within the single time step of their arrival. Each consumer has a flexibility level, i.e., a number between $1$ and $k$ that indicates the varieties of the goods the consumer finds equally desirable. A consumer's    flexibility level as well as the utility it enjoys upon allocation of a desired good  are its private information.  We characterized the allocation and payment functions under the optimal mechanism in terms of the solution to a dynamic program. 
We leveraged the structure of the consumers' flexibility model to simplify the dynamic program and provided an alternative description of the optimal mechanism in terms of thresholds computed by the dynamic program.

An interesting extension to the present work would be to study this setup with \textit{patient} consumers, i.e., consumers may be present for more than one time step. In addition, studying the dynamic mechanism design problem under the settings where both arrival and departure times of each consumer are  privately known to them is an important  direction for further exploration. In the present setup we studied the case where each consumer wants to receive a \textit{single} good of its desired varieties. Another interesting scenario would be the case where the consumers may need to get \textit{multiple} goods of their desired varieties. 

\appendices

\section{Proof of Lemma \ref{lem:revEqBern}} \label{sec:revEqBern-pf}
Consider a BIC and IR mechanism $(q_{1:T}, p_{1:T})$. The expected revenue under this mechanism  is
\begin{align}
&\mathbb{E}\Big\{ \sum\limits_{t=1}^T  \sum\limits_{i =1}^{N_t} p^{i}_{t}(H_{t}) \Big\} \notag \\
&=\sum\limits_{t=1}^T \sum\limits_{n_t=1}^{\bar{n}} \lambda_t(n_t)  \;  \sum\limits_{i =1}^{n_t} \mathbb{E}\Big[  p^{i}_{t}(H_{t}) \; | \; N_t = n_t \Big].    \label{eq:ExRev}
\end{align}
The conditional expectation in (\ref{eq:ExRev}) can be written as
\begin{align}
&\sum\limits_{\tilde{b}_t^i=1}^k \int\limits_{\theta^{\text{min}}}^{\theta^{\text{max}}}  \mathbb{E}_{H^{-i}_t}[\; p^{i}_{t}(H^{-i}_{t}, (\tilde{\theta}_t^i, \tilde{b}_t^i)) \; | N_t = n_t  \; ]  \; f_t(\tilde{\theta}_t^i, \tilde{b}_t^i) \; d\tilde{\theta}^i_t \; \notag \\
&= \sum\limits_{\tilde{b}_t^i=1}^k \int\limits_{\theta^{\text{min}}}^{\theta^{\text{max}}}  P^{i}_t(\tilde{\theta}_t^i, \tilde{b}_t^i, n_t)  \; f_t(\tilde{\theta}_t^i, \tilde{b}_t^i) \; d\tilde{\theta}^i_t \label{eq:sumP}
\end{align}
where $P^{i}_t(\cdot)$ is the interim payment defined in (\ref{eq:P}). Because of Lemma \ref{lem:BICthetaIR}, we know that
\begin{align}
\hspace{-1em}P^i_{t}( \tilde{\theta}_t^i, \tilde{b}_t^i, n_t) &\le \tilde{\theta}_t^i \; Q^i_{t}( \tilde{\theta}_t^i, \tilde{b}_t^i, n_t) - \int\limits_{\theta^{\text{min}}}^{\tilde{\theta}_t^i} Q^i_{t}(\alpha, \tilde{b}_t^i, n_t) \; d\alpha. \label{eq:stepP}
\raisetag{1.8\baselineskip}
\end{align}
Using (\ref{eq:stepP}) and after simplifying the integrals, (\ref{eq:sumP}) can be upper bounded by the following expression
\begin{align}\label{eq:sumQw}
\sum\limits_{\tilde{b}_t^i=1}^k \;   \int\limits_{\theta^{\text{min}}}^{\theta^{\text{max}}}    \; Q^{i}_t(\tilde{\theta}_t^i, \tilde{b}_t^i,  n_t) \; w_t(\tilde{\theta}_t^i, \tilde{b}_t^i)  f_t(\tilde{\theta}_t^i, \tilde{b}_t^i) \; d\tilde{\theta}_t^i,
\end{align}
The upper bound in (\ref{eq:sumQw})  implies that  the expected total revenue  in (\ref{eq:ExRev}) can be upper bounded by the following 
\begin{align}
&\sum\limits_{t=1}^T \sum\limits_{n_t=1}^{\bar{n}} \lambda_t(n_t)  \;  \sum\limits_{i =1}^{n_t} \sum\limits_{\tilde{b}_t^i=1}^k \;  \int\limits_{\theta^{\text{min}}}^{\theta^{\text{max}}}    \notag \\
&\mathbb{E}\Big[ \; w_t(\tilde{\theta}_t^i, \tilde{b}_t^i) \sum\limits_{j \le \tilde{b}_t^i} q_{t}^{i, j}(H^{-i}_{t}, (\tilde{\theta}_t^i, \tilde{b}_t^i))\;  | \; N_t = n_t \Big]  
\; f_t(\tilde{\theta}_t^i, \tilde{b}_t^i ) \; d\tilde{\theta}_t^i \; , \notag \\
&= \sum\limits_{t=1}^T \sum\limits_{n_t=1}^{\bar{n}} \lambda_t(n_t)  \; \sum\limits_{i=1}^{n_t}   \mathbb{E}\Big[ \; w_t(\theta^i_t, b^i_t) \sum\limits_{j \le b_t^i} q_{t}^{i, j}(H_{t})\;   | \; N_t = n_t \Big]  \notag \\
& = \sum\limits_{t=1}^T   \mathbb{E}\Big[ \;  \sum\limits_{i=1}^{N_t}  w_t(\theta^i_t, b^i_t) \sum\limits_{j \le b^i_t} q_{t}^{i, j}(H_{t})\;    \Big]  \notag \\
& = \mathbb{E}\Big[ \;  \sum\limits_{t=1}^T   \sum\limits_{i=1}^{N_t}  w_t(\theta^i_t, b^i_t) \sum\limits_{j \le b^i_t} q_{t}^{i, j}(H_{t})\; \Big] \notag \\
&\le  \max\limits_{q_{1:T}} \; \mathbb{E}\Big[ \;   \sum\limits_{t=1}^T   \sum\limits_{i=1}^{N_t} w_t(\theta_t^i, b_t^i) \; \Big(  \sum\limits_{j \le b_t^i} q_{t}^{i, j}(H_{t}) \Big)  \; \Big]. \notag
\end{align}
Thus, the expected revenue of any BIC and IR mechanism is upper bounded by the maximum value in \eqref{revenue}. Consequently, a BIC and IR mechanism $(q^*_{1:T}, p^*_{1:T})$ for which $q^*_{1:T}$  achieves the maximum value in \eqref{revenue} and $p^*_{1:T}$  are of the form given in (\ref{eq:expostp}),\footnote{$p^*_{1:T}$  form in (\ref{eq:expostp}) make  the  upper bound on the expected total revenue  attainable, by ensuring that the inequality in (\ref{eq:stepP}) becomes an equality for $(q^*_{1:T}, p^*_{1:T})$.}   guarantees the largest expected revenue among all BIC and IR mechanisms. This concludes the proof.

\section{Proof of Lemma \ref{lem:infoState}}\label{sec:infoStatePf}
We prove this by induction.

\textit{Base case:}  Clearly the expression given for $R_T(h_T)$ in (\ref{eq:DPqtempty})-(\ref{eq:DPqt}) solely depends on $h^R_T$ and $y_T^{1:k}$ (recall that $R_{T+1}(\cdot) =0$). That is, the information in $h_T \setminus \{ h^R_T, y_T^{1:k} \}$ is irrelevant for determining  $R_T(h_T)$. Therefore, if we define the function $V_T(\cdot)$ as:
\begin{itemize}
\item If $h^R_T = \emptyset \;\; : \;\; V_T(h^R_T, y_T^{1:k}) := 0.$
\item If $h^R_T \neq \emptyset \;\; :$
\begin{align*}
\hspace{-2em}V_T(h^R_T, y_T^{1:k}) := \max\limits_{\bold{A} \in \mathcal{S}(h^R_T, y_T^{1:k})} \Big\{ \; \sum\limits_{i =1}^{|h^R_T|} w_T( \tilde{\theta}_T^i, \tilde{b}_T^i) \sum\limits_{j=1}^k \bold{A}(i,j) \Big\},
\end{align*}
\end{itemize}
the equality in (\ref{eq:Vfuncs}) holds true at time $T$.

\textit{Induction hypothesis:} Suppose there exists some  function $V_{t+1}(\cdot)$ such that (\ref{eq:Vfuncs}) holds true at time $t+1$. 

Now, we want to show that there exists some function $V_t(\cdot)$ such that (\ref{eq:Vfuncs}) holds true at time $t$. In other words, we want to show that given the history $h_t = \{ h^R_t, y_t^{1:k}, x_t^{1:k}, h_{t-1} \}$, the expression given for $R_t(\cdot)$ in  (\ref{eq:DPqtempty})-(\ref{eq:DPqt}) is fully determined from  $\{ h^R_t, y_t^{1:k} \}$ and does not depend on $h_t \setminus \{ h^R_t, y_t^{1:k} \} = \{ x_t^{1:k}, h_{t-1} \}$. For the case $h^R_t = \emptyset$, from (\ref{eq:DPqtempty}) we see that  $R_t(h_t)$ is expressed as:
\begin{align*}
R_t(h_t) &=  \mathbb{E}\Big[R_{t+1}(h_t, H^R_{t+1}, \{y_t^j+X_{t+1}^j\}_{j=1}^k, X_{t+1}^{1:k})  \Big].
\end{align*}
Using the induction hypothesis, the above expression can be written as:
\begin{align*}
R_t(h_t) &=  \mathbb{E}\Big[V_{t+1}(H^R_{t+1}, \{y_t^j+X_{t+1}^j\}_{j=1}^k)  \Big].
\end{align*}
Since $H^R_{t+1}$ and $X^j_{t+1}$ are independent of $h_t$, the expected value above depends only on $y^{1:k}_t$. Thus, when $h^R_t=\emptyset$, we can define
\begin{align*}
V_t(\emptyset, y^{1:k}_t) = R_t(h_t) = \mathbb{E}_{H^R_{t+1},Y^{1:k}_{t+1}}\Big[V_{t+1}(H^R_{t+1}, Y^{1:k}_{t+1}) \Big].
\end{align*} 
Note that the above definition of $V_t(\cdot)$ satisfies (\ref{eq:DPVqtempty}).

For the case $h^R_t \neq \emptyset$, we see from (\ref{eq:DPqt})  that  $R_t(h_t)$ is expressed as:
\begin{align}
R_t(h_t) &= \max\limits_{\bold{A} \in \mathcal{S}(h^R_t, y_t^{1:k})} \Big\{ \;  \underbrace{\sum\limits_{i =1}^{|h^R_t|} w_t(  \tilde{\theta}_t^i, \tilde{b}_t^i) \sum\limits_{j=1}^k \bold{A}(i,j)}_{\dagger} \notag \\
&+ \mathbb{E}\Big[R_{t+1}(H_{t+1}) | h_t, \bold{A}_t = \bold{A} \Big] \Big\}. \label{eq:Rtstep*}
\end{align}
Clearly, the term $\dagger$ in the above expression  does not depend on the information in $h_t \setminus \{ h^R_t, y_t^{1:k} \} = \{ x_t^{1:k}, h_{t-1} \}$.  Moreover the set $\mathcal{S}(h^R_t, y_t^{1:k})$ over whose elements the $\max\{ \cdot \}$ operation above is carried out is fully specified  in terms of $h^R_t, y_t^{1:k}$ and does not depend on $\{ x_t^{1:k}, h_{t-1} \}$. It thus remains to show that the second term in the $\max\{ \cdot \}$ operation above, does not depend on the information in $\{ x_t^{1:k}, h_{t-1} \}$ either.  Using the induction hypothesis, let us expand the second term above as follows:
\begin{align}
&\mathbb{E}\Big[R_{t+1}(H_{t+1}) | H_t = h_t, \bold{A}_t = \bold{A} \Big] \notag \\ 
&= \mathbb{E}\Big[ V_{t+1}(H_{t+1}^R, Y_{t+1}^{1:k}) |  H_t = h_t, \bold{A}_t = \bold{A} \Big] \notag \\
&= \mathbb{E}\Big[ V_{t+1}(H_{t+1}^R, \notag \\
&\{y_t^j - \sum\limits_{i=1}^{|h^R_t|}\bold{A}(i,j) + X_{t+1}^j\}_{j=1}^k) \; |\; \{ h^R_t, y_t^{1:k}, x_t^{1:k}, h_{t-1} \}, \bold{A}\Big]  \notag \\
&= \mathbb{E}\Big[ V_{t+1}(H_{t+1}^R, \{y_t^j - \sum\limits_{i=1}^{|h^R_t|}\bold{A}(i,j) + X_{t+1}^j\}_{j=1}^k) \; |\; h^R_t, y_t^{1:k}, \bold{A}\Big]  \notag \\
&= \mathbb{E}\Big[  V_{t+1}(H_{t+1}^R, Y_{t+1}^{1:k}) \;|\; h^R_t, y_t^{1:k}, \bold{A} \Big], \label{eq:Rtstep**}
\end{align}
where we used the fact that $H^R_{t+1}$ and $X^{1:k}_{t+1}$ are independent of $h_t$.
It is clear that the above conditional expectation is a function of $h^R_t, y_t^{1:k}, \bold{A}$ and does not depend on the information in $\{ x_t^{1:k}, h_{t-1} \}$. 
The above analysis allows us to conclude that 
\begin{enumerate}[(i)]
\item $R_t(h_t)$ is completely determined by $h^R_t , y^{1:k}_t$. Thus, we can define a function   $V_t(h^R_t , y^{1:k}_t) = R_t(h_t)$.
\item Further, using (\ref{eq:Rtstep*}) and (\ref{eq:Rtstep**}) above, it is clear that $V_t(\cdot)$ satisfies (\ref{eq:DPVqt}).
\end{enumerate}
 This completes the proof.

%

\section{Proof of Lemma \ref{lem:Uset}}\label{sec:UsetPf}

 Let $\mathcal{A} := \Big\{  u^{1:k} \in \mathbb{Z}^k_{\ge 0} :  \sum\limits_{l=1}^j u^l \le \sum\limits_{l=1}^j y_t^l \; , \; u^j \le n^{j}_t, \forall j  \Big\}$, i.e., $\mathcal{A}$ equals the set in the right-hand side of (\ref{eq:Uset}). Clearly when $h^R_t = \emptyset$, $\mathcal{U}(h^R_t, y_t^{1:k})  := \{ \bold{0}_{1 \times k} \} = \mathcal{A}$. 
 
 Let us now consider the information state $s_t = (h^R_t, y_t^{1:k})$ with $h^R_t \neq \emptyset$. We start with showing that $\mathcal{U}(h^R_t, y_t^{1:k}) \subseteq \mathcal{A}$.
Consider a vector $u_t^{1:k} \in \mathcal{U}(h^R_t, y_t^{1:k})$. 
Since $n^j_t$ is the number of consumers with flexibility level $j$ that arrive at time $t$ and  $u^j_t$ is the number of consumers with flexibility level $j$ that get a good at time $t$, we clearly have that $u^j_t \leq n^j_t$.

Now, consider $\sum_{l=1}^j u_t^l$. This is the total number of consumers with flexibility level less than or equal to $j$ that get a good. Since consumers cannot get a good of variety higher than their flexibility level, it follows that  $\sum_{l=1}^j u_t^l$ is less than or equal to the total number of available goods of variety less than or equal to $j$. In other words,
$\sum_{l=1}^j u_t^l \leq \sum_{l=1}^j y_t^l$. Thus, $u_t^{1:k} \in \mathcal{A}$.


We now show that script $\mathcal{A} \subseteq \mathcal{U}(h^R_t, y_t^{1:k})$. Consider $u_t^{1:k} \in \mathcal{A}$. 
In order to prove that $u_t^{1:k} \in  \mathcal{U}(h^R_t, y_t^{1:k})$, we need to show that there exists some matrix $\bold{D} \in  \mathcal{S}(h^R_t, y_t^{1:k})$ such that $\sum\limits_{\substack{i=1 \\ i: \tilde{b}^i_t = j}}^{|h^R_t|} \sum\limits_{l \le j} \bold{D}(i,l) = u_t^j, \forall j$. Let us construct such a matrix according to the following allocation procedure:
\begin{itemize}
\item Select any $u_t^1$ consumers with flexibility level 1 and allocate each of them a good of variety 1. This is a feasible allocation since $u_t^1 \leq y^1_t$.
\item Select any $u_t^2$ consumers with flexibility level 2 and allocate each of them either an unallocated good of  variety 1 (if $u_t^1 < y_t^1$) or a good of variety 2. These can be done since $u_t^1 + u_t^2 \le y^1_t + y_t^2$. 
\item Proceed in a similar fashion for all flexibility levels: select any $u_t^j$ consumers with flexibility level $j$ and allocate each of them a good of any of the varieties $1, \ldots, j$ depending on their availability. Since $\sum\limits_{l=1}^j u_t^l \le \sum\limits_{l=1}^j y_t^l$   the described allocation is feasible.
\item The other consumers that arrived at time $t$ but were not selected for allocation in the above steps get zero allocation. 
\end{itemize}
It is straightforward to verify that allocation matrix $\bold{D}$ constructed above  belongs to $\mathcal{S}(h^R_t, y_t^{1:k})$ and that it serves $u_t^j$ consumers of flexibility level $j$.
Hence, every vector $u_t^{1:k} \in \mathcal{A}$  corresponds to a feasible allocation matrix $\bold{D} \in  \mathcal{S}(h^R_t, y_t^{1:k})$. Hence   $u_t^{1:k} \in \mathcal{U}(h^R_t, y_t^{1:k})$. This establishes $\mathcal{A} \subseteq \mathcal{U}(h^R_t, y_t^{1:k})$ and completes the proof.



\section{Proof of Lemma \ref{lem:DPuv}} \label{sec:DPuvPf}
Define
\begin{equation}\label{F}
\begin{split}
\mathcal{F}(h^R_t, y_t^{1:k}) := \Big\{  &(u^{1:k}, v^{1:k}) \in \mathbb{Z}_{\ge 0}^{2k} : \\
&u^{1:k} \in \mathcal{U}(h^R_t, y_t^{1:k}) \; , \; v^{1:k} \in \mathcal{V}(u^{1:k},y_t^{1:k})   \Big\},
\end{split}
\raisetag{2.5\baselineskip}
\end{equation}
Further, for any $(u^{1:k} , v^{1:k})$ in $\mathcal{F}(h^R_t, y_t^{1:k})$ define
\begin{equation}\label{SF}
\begin{split}
&\mathcal{S}_{\mathcal{F}}(h^R_t, y_t^{1:k}, u^{1:k}, v^{1:k}) := \Big\{ \bold{A} \in \mathcal{S}(h^R_t, y_t^{1:k}) : \\
&\sum\limits_{i=1}^{|h^R_t|} \bold{A}(i,j) = v^j,  \sum\limits_{\substack{i = 1 \\ i: \tilde{b}_t^i = j}}^{|h^R_t|} \sum\limits_{l=1}^j \bold{A}(i,l) = u^j \; , j = 1, \ldots, k  \Big\}.
\end{split}
\raisetag{4\baselineskip}
\end{equation}
It is easy to check that the set of all feasible allocation matrices can be partitioned as
\begin{equation}\label{FSeq}
\mathcal{S}(h^R_t, y_t^{1:k}) = \bigcup\limits_{(u^{1:k}, v^{1:k}) \in \mathcal{F}(h^R_t, y_t^{1:k})} \mathcal{S}_{\mathcal{F}}(h^R_t, y_t^{1:k}, u^{1:k}, v^{1:k}).
\end{equation}
Therefore, the value function in (\ref{eq:DPVqt}) can be written as
\begin{align}
&V_t(s_t) = \max\limits_{(u^{1:k}, v^{1:k}) \in \mathcal{F}(s_t)} \Big\{ \;  \max\limits_{\bold{A} \in \mathcal{S}_{\mathcal{F}}(s_t, u^{1:k}, v^{1:k})} \Big\{ \; \notag \\
&\sum\limits_{i =1}^{|h^R_t|} w_t(  \tilde{\theta}_t^i, \tilde{b}_t^i) \sum\limits_{j=1}^k \bold{A}(i,j) + \underbrace{\mathbb{E}\Big[V_{t+1}(S_{t+1}) | s_t, \bold{A}_t = \bold{A} \Big]}_{\dagger}  \Big\} \Big\}, \label{eq:future}
\raisetag{4\baselineskip}
\end{align}
The  $\dagger$  term in (\ref{eq:future}) can be written as
\begin{align}
\dagger = \mathbb{E}\Big[ V_{t+1}\Big( H^R_{t+1} , \{y^j_t - v^j + X^j_{t+1} \}_{j=1}^k \Big) \Big] \notag
\end{align}
The above expectation depends only on $v^{1:k}$ and $y^{1:k}_t$ and not on the allocation matrix itself. Thus (\ref{eq:future}) becomes
\begin{align}
&V_t(s_t) = \notag \\
&\max\limits_{(u^{1:k}, v^{1:k}) \in \mathcal{F}(s_t)} \Big\{  \mathbb{E}\Big[V_{t+1}\Big( H^R_{t+1} , \{y^j_t - v^j + X^j_{t+1} \}_{j=1}^k \Big) \Big] \notag \\
&+\underbrace{\max\limits_{\bold{A} \in \mathcal{S}_{\mathcal{F}}(s_t, u^{1:k}, v^{1:k})} \Big\{ \;\sum\limits_{i =1}^{|h^R_t|} w_t(  \tilde{\theta}_t^i, \tilde{b}_t^i) \sum\limits_{j=1}^k \bold{A}(i,j)  \Big\}}_{\ddagger} \Big\}, \label{eq:maxWs_t}
\end{align}
Considering the term $\ddagger$ in the above expression, it is straightforward to see that for all $\bold{A} \in \mathcal{S}_{\mathcal{F}}(s_t , u^{1:k}, v^{1:k})$,
\begin{align*}
\sum\limits_{i =1}^{|h^R_t|} w_t(  \tilde{\theta}_t^i, \tilde{b}_t^i) \sum\limits_{j=1}^k \bold{A}(i,j) \le \sum\limits_{j=1}^k\sum\limits_{i=1}^{u^j} w_t^{i,j},
\end{align*}
where $w_t^{i,j}$ denotes the $i$th largest element in $\mathcal{W}_t^j$ (see (\ref{eq:Wjt})). Further, an allocation matrix that, for each flexibility level $j$,  gives goods to $u^j$ consumers with highest virtual valuations satisfies the above inequality with equality. Hence, 
 \begin{align*}
&\max\limits_{\bold{A} \in \mathcal{S}_{\mathcal{F}}(s_t, u^{1:k}, v^{1:k})} \Big\{ \;\sum\limits_{i =1}^{|h^R_t|} w_t(  \tilde{\theta}_t^i, \tilde{b}_t^i) \sum\limits_{j=1}^k \bold{A}(i,j)  \Big\} = \sum\limits_{j=1}^k\sum\limits_{i=1}^{u^j} w_t^{i,j}.
\end{align*}
Plugging this result into (\ref{eq:maxWs_t}) we obtain
\begin{align}
&V_t(h^R_t, y_t^{1:k}) = \max\limits_{(u^{1:k}, v^{1:k}) \in \mathcal{F}(h^R_t, y_t^{1:k})} \Big\{ \;  \sum\limits_{j=1}^k\sum\limits_{i=1}^{u^j} w_t^{i,j} \notag \\
&+\mathbb{E}\Big[V_{t+1}\Big( H^R_{t+1} , \{y^j_t - v^j + X^j_{t+1} \}_{j=1}^k \Big) \Big]  \Big\}, \label{eq:VtF}
\end{align}
which can be rearranged in the form of the  nested maximizations  below
\begin{align}
&V_t(h^R_t, y_t^{1:k}) = \max\limits_{u^{1:k} \in \mathcal{U}(h^R_t, y_t^{1:k})} \Big\{ \;  \sum\limits_{j=1}^k\sum\limits_{i=1}^{u^j} w_t^{i,j} \notag \\
&+ \max\limits_{v^{1:k} \in \mathcal{V}(u^{1:k}, y_t^{1:k})} \Big\{ \; \mathbb{E}\Big[V_{t+1}\Big( H^R_{t+1} , \{y^j_t - v^j + X^j_{t+1} \}_{j=1}^k \Big) \Big] \;  \Big\}  \Big\} \label{eq:uvsplit}
\raisetag{3\baselineskip}
\end{align}
This completes the proof.

\section{Proof of Lemma \ref{lem:Rmonotone}} \label{sec:yzPf}
We  provide an inductive proof of the lemma.

\textit{Base Case:} At time  $T$, consider a non-empty history $h^R_T$ and supply profiles $y_T^{1:k}$ and $z_T^{1:k}$ such that $y^i_T = z_T^i + 1 \; , \; y^j_T = z_T^j - 1$ and $y^l_T = z_T^l$ for $l \neq i,j$, where $i<j$.
From the definition of $\mathcal{U}( \cdot)$ in (\ref{eq:Uset}) it follows  that $\mathcal{U}(h^R_T, z_T^{1:k}) \subseteq \mathcal{U}(h^R_T, y_T^{1:k})$.  This fact combined with the definition of $V_T(\cdot)$ implies that (\ref{eq:def1Rmon}) holds for $t = T$ and $h^R_T \neq \emptyset$. 
If $h^R_T = \emptyset$, then $V_T(h^R_T, z_T^{1:k}) = 0 = V_T(h^R_T, y_T^{1:k})$.
Hence (\ref{eq:def1Rmon}) holds true at time $T$ for all $h^R_T$.

\textit{Induction hypothesis:}  Suppose that the statement of the lemma is true for $V_{t+1}(\cdot)$. Consider two supply profiles $y_{t}^{1:k}$ and $z_{t}^{1:k}$ such that $y^i_{t} = z_{t}^i + 1 \; , \; y^j_{t} = z_{t}^j - 1$ and $y^l_{t} = z_{t}^l$ for $l \neq i,j$, where $i<j$.
We now  show that given such $y_{t}^{1:k}$ and $z_{t}^{1:k}$ the property in (\ref{eq:def1Rmon}) holds true at time $t$, i.e., 
\begin{align}
V_{t}(h^R_{t}, y_{t}^{1:k})  \ge V_{t}(h^R_{t}, z_{t}^{1:k}) \; , \; \forall h^R_{t}. \label{eq:Vatt}
\end{align}
Let us first consider $h^R_t = \emptyset$. In this case, we have
\begin{align}
V_t(\emptyset, y_t^{1:k}) &= \mathbb{E}\Big[ \; V_{t+1}(H^R_{t+1}, Y_{t+1}^{1:k})    \; | \;  y_t^{1:k} \;\Big] \notag \\
&= \mathbb{E}\Big[ \; V_{t+1}(H^R_{t+1}, \{ y_t^l+X_{t+1}^l \}_{l=1}^k)  \;\Big] \notag \\
&\ge \mathbb{E}\Big[ \; V_{t+1}(H^R_{t+1}, \{ z_t^l+X_{t+1}^l \}_{l=1}^k)   \;\Big] \label{eq:useIHyz2} \\
&= \mathbb{E}\Big[ \; V_{t+1}(H^R_{t+1}, Y_{t+1}^{1:k})   \; | \;  z_t^{1:k}  \;\Big] \notag \\
&= V_t(\emptyset, z_t^{1:k}), \notag
\end{align}
where the  inequality in (\ref{eq:useIHyz2}) follows from the induction hypothesis   by noting that  for any  realization $x^{1:k}_{t+1}$,  $y^i_{t+1} = y^i_t + x^i_{t+1} = z_{t+1}^i + 1 \; , \; y^j_{t+1} = y^j_t + x^j_{t+1} = z_{t+1}^j - 1$ and $y^l_{t+1} =  y^l_t + x^l_{t+1}  = z_{t+1}^l$ for $l \neq i,j$. This establishes the property in (\ref{eq:def1Rmon}) for $h^R_t = \emptyset$.

Now consider $h^R_t \neq \emptyset$. To prove (\ref{eq:def1Rmon}), it suffices to show that for every $(u^{1:k},v^{1:k}) \in \mathcal{F}(h^R_t, z_t^{1:k})$, there exists $(u^{1:k},a^{1:k}) \in \mathcal{F}(h^R_t, y_t^{1:k})$ such that 
\begin{align}
&\sum\limits_{j=1}^k\sum\limits_{i=1}^{u^j} w_t^{i,j} +\mathbb{E}\Big[V_{t+1}(H^R_{t+1}, \{ y_t^l - a^l +X_{t+1}^l \}_{l=1}^k) \Big] \notag \\ 
&\ge \sum\limits_{j=1}^k\sum\limits_{i=1}^{u^j} w_t^{i,j}  +\mathbb{E}\Big[V_{t+1}(H^R_{t+1}, \{ z_t^l - v^l +X_{t+1}^l \}_{l=1}^k) \Big] \label{eq:revToGoDom}
\raisetag{4\baselineskip}
\end{align}
Recall that $(u^{1:k},v^{1:k}) \in \mathcal{F}(h^R_t, z_t^{1:k})$ implies that $v^l \le z^l_t$ for all $l$.

For $(u^{1:k},v^{1:k}) \in \mathcal{F}(h^R_t, z_t^{1:k})$ two cases arise:
\begin{itemize}
\item Case 1: $v^j < z^j_t$.

In this case we define $a^{1:k} = v^{1:k}$. It is clear that $(u^{1:k},a^{1:k}) \in \mathcal{F}(h^R_t, y_t^{1:k})$ and (\ref{eq:revToGoDom}) holds.

\item Case 2: $v^j =  z^j_t$.

In this case we cannot set $a^{1:k} = v^{1:k}$ since $v^j > y^j_t$. Therefore, we define $a^{1:k}$ as follows: $a^i = v^i + 1, a^j = v^j - 1$ and $a^l = v^l$ for $l \neq i,j$. It is straightforward to verify that $(u^{1:k},a^{1:k}) \in \mathcal{F}(h^R_t, y_t^{1:k})$. Further, using the induction hypothesis 
\begin{align}
&\mathbb{E}\Big[V_{t+1}(H^R_{t+1}, \{ y_t^l - a^l +X_{t+1}^l \}_{l=1}^k) \Big] \notag \\
&= \mathbb{E}\Big[V_{t+1}(H^R_{t+1}, y_t^1 - a^1 +X_{t+1}^1, \ldots, y_t^i - a^i +X_{t+1}^i, \notag \\
&\ldots, y_t^j - a^j+X_{t+1}^j, \ldots, y_t^k - a^k +X_{t+1}^k) \Big] \notag \\
&= \mathbb{E}\Big[V_{t+1}(H^R_{t+1}, y_t^1 - v^1 +X_{t+1}^1, \ldots \notag \\
&\ldots, y_t^i - (v^i+1) +X_{t+1}^i, \ldots, y_t^j - (v^j-1)+X_{t+1}^j, \ldots \notag \\
& \ldots, y_t^k - v^k +X_{t+1}^k) \Big] \notag\\
&= \mathbb{E}\Big[V_{t+1}(H^R_{t+1}, y_t^1 - v^1 +X_{t+1}^1, \ldots \notag \\
&\ldots, (y_t^i-1) - v^i +X_{t+1}^i, \ldots, (y_t^j+1) - v^j +X_{t+1}^j, \ldots \notag \\
&\ldots , y_t^k - v^k +X_{t+1}^k) \Big] \notag \\
&= \mathbb{E}\Big[V_{t+1}(H^R_{t+1}, \{ z_t^l - v^l +X_{t+1}^l \}_{l=1}^k) \Big]. \notag
\end{align}
This proves (\ref{eq:revToGoDom}) and thus, establishes (\ref{eq:def1Rmon}) for $h^R_t \neq \emptyset$
\end{itemize}

Therefore, (\ref{eq:def1Rmon}) holds true for all $h^R_t$ at time $t$. This completes the proof.

\section{Proof of Lemma \ref{lem:v*recipe}}\label{sec:v*recipePf}
We prove the lemma in the following steps:

\textbf{Step 1:} We first show that $v^{*1:k} \in \mathcal{V}(u^{1:k}, y_t^{1:k})$. Clearly $v^{*j} \le y^j_t$ and $\sum_{l=j}^k v^{*l} \le \sum_{l=j}^k u^l$ for all $j$. To show that $v^{*1:k} \in \mathcal{V}(u^{1:k}, y_t^{1:k})$, it suffices to show that $\sum_{l=1}^k v^{*l} = \sum_{l=1}^k u^l$. We will utilize the following claim.

\textit{Claim:} Suppose that for all $m = 1, \ldots, j$: (i) $\sum_{l=m}^k v^{*l} < \sum_{l=m}^k u^l$ and, (ii) $v^{*i} = y^i_t$, for all $i < j$. Then, (a) $v^{*j} = y^j_t$ and, (b) $\sum_{l=j+1}^k v^{*l} < \sum_{l=j+1}^k u^l$.

Assume for now that the claim is true.  We have already seen that $\sum_{l=1}^k v^{*l} \le \sum_{l=1}^k u^l$. Suppose 
\begin{equation}\label{Assumvlessu}
\sum\limits_{l=1}^k v^{*l} < \sum\limits_{l=1}^k u^l.
\end{equation} 

Then, using the claim above with $j=1$ implies that  $v^{*1} = y^1_t$ and $\sum_{l=2}^k v^{*l} < \sum_{l=2}^k u^l$. We can now use the above claim for $j=2$ to conclude that $v^{*2} = y^2_t$ and $\sum_{l=3}^k v^{*l} < \sum_{l=3}^k u^l$. Proceeding this way until $j=k-1$, we get that (1) $v^{*m} = y_t^m$ for all $m \le k-1$  and (2) $v^{*k} < u^k$. Further, (2) and definition of $v^{*k}$ implies that $v^{*k}=y_t^k$. Thus, the entire $v^{*1:k}$ vector is equal to the $y^{1:k}_t$ vector. But we started with the statement that equation \eqref{Assumvlessu} is true. Thus, $\sum_{l=1}^k y_t^{l} < \sum_{l=1}^k u^l$ which contradicts the fact that $u^{1:k} \in \mathcal{U}(h^R_t, y_t^{1:k})$. Thus, equation \eqref{Assumvlessu} is false and hence $\sum\limits_{l=1}^k v^{*l} = \sum\limits_{l=1}^k u^l$.

The only thing left now is the proof of the claim. 

\emph{Proof of claim:} The inequality  $\sum_{l=j}^k v^{*l} < \sum_{l=j}^k u^l$   implies that $v^{*j} < \sum_{l=j}^k u^l  - \sum_{l=j+1}^k v^{*l}$. This, along with the definition of $v^{*j}$ implies that $v^{*j} = y^j_t$. Therefore (a) holds true. 

We already know that $\sum_{l=j+1}^k v^{*l} \le \sum_{l=j+1}^k u^l$. If $\sum_{l=j+1}^k v^{*l} = \sum_{l=j+1}^k u^l$, then from (i) in the claim statement with $m=1$ it follows that $\sum_{l=1}^j v^{*l} < \sum_{l=1}^j u^l$. This combined with $v^{*i} = y^i_t$, for all $i < j$ in (ii) and  $v^{*j} = y^j_t$ in part (a)  implies that $\sum_{l=1}^j y^l_t < \sum_{l=1}^j u^l$ which contradicts the fact that $u^{1:k} \in \mathcal{U}(h^R_t, y_t^{1:k})$. Hence, $\sum_{l=j+1}^k v^{*l} = \sum_{l=j+1}^k u^l$ cannot be true. This establishes (b). \\

\textbf{Step 2:} For any $v^{1:k} \neq v^{*1:k}$ in $\mathcal{V}(u^{1:k}, y_t^{1:k})$ consider the highest $j$ such that $v^j \neq v^{*j}$. We argue that  $v^j > v^{*j}$ cannot be true.  Given the definition of $v^{*j}$,  either $v^{*j} = y^j_t$ or $v^{*j} = u^j + (\sum_{l=j+1}^k u^l - \sum_{l=j+1}^k v^{*l} )$. If $v^{*j} = y^j_t$, then $v^j > v^{*j}$ contradicts $v^{1:k} \in \mathcal{V}(u^{1:k}, y_t^{1:k})$. Now suppose $v^{*j} = u^j + (\sum_{l=j+1}^k u^l - \sum_{l=j+1}^k v^{*l})$. Given that $v^l = v^{*l}, l = j+1, \ldots, k$, $v^j > v^{*j}$ would then imply that $v^j > u^j + (\sum_{l=j+1}^k u^l - \sum_{l=j+1}^k v^{l})$ or arranged differently $\sum_{l=j}^k v^{l} > \sum_{l=j}^k u^l$. This combined with the fact that $\sum_{l=1}^k v^l = \sum_{l=1}^k u^l$, would then imply that $\sum_{l=1}^{j-1} v^l < \sum_{l=1}^{j-1} u^l$, which contradicts $v^{1:k} \in \mathcal{V}(u^{1:k}, y_t^{1:k})$. Thus it can only be the case that $v^j < v^{*j}$. \\

\textbf{Step 3:} For any $v^{1:k} \neq v^{*1:k}$ in $\mathcal{V}(u^{1:k}, y_t^{1:k})$, we define a new vector $\mathcal{T}(v^{1:k})$ as follows: Pick the highest $j$ such that $v^j < v^{*j}$.\footnote{Note that the case $v^j > v^{*j}$ got ruled out in Step 2.} Then, pick the highest $i<j$ with $v^i > 0$. It can easily be shown that such $i$ and $j$ exist. Then, $\mathcal{T}^j(v^{1:k})=v^j+1, \mathcal{T}^i(v^{1:k})=v^i-1$ and $\mathcal{T}^l(v^{1:k}) = v^l$ for $l \neq i , j$, where $\mathcal{T}^l(v^{1:k})$ denotes the $l$th entry in $\mathcal{T}(v^{1:k})$.  We now argue that  $\mathcal{T}(v^{1:k}) \in \mathcal{V}(u^{1:k}, y_t^{1:k})$.  For  $m<i$ or $m \geq j$, it is clear that
\[    \sum_{l=1}^m u^l  \leq  \sum_{l=1}^m \mathcal{T}^l(v^{1:k}).    \]
Since $j$ is the highest index with $v^j \neq v^{*j}$, it follows that 
\[   \sum_{l=1}^{j-1} v^l > \sum_{l=1}^{j-1} v^{*l} \ge \sum_{l=1}^{j-1} u^l.  \]
Now, for any $m$ such that $i \leq m < j$,
\[   \sum_{l=1}^m \mathcal{T}^l(v^{1:k}) = (v^i - 1) + \sum_{l=1}^{i-1} v^l  = \Big( \sum_{l=1}^{j-1} v^l \Big) - 1 \geq \sum_{l=1}^m u^l.\]
Therefore $\mathcal{T}(v^{1:k})$ satisfies all the  inequalities $\sum_{l=1}^{m} \mathcal{T}^l(v^{1:k}) \ge \sum_{l=1}^{m} u^l, m = 1, \ldots, k$. Further, it is easy to verify that $\mathcal{T}^j(v^{1:k}) \le y_t^j$. Thus, $\mathcal{T}(v^{1:k}) \in \mathcal{V}(u^{1:k}, y_t^{1:k})$.
We now show that  the objective value in (\ref{eq:v*}) is (weakly) larger under $\mathcal{T}(v^{1:k})$ compared to that under $v^{1:k}$. Let $a^{1:k} := \mathcal{T}(v^{1:k})$.  Using Lemma \ref{lem:Rmonotone} 
\begin{align}
&\mathbb{E}[V_{t+1}(H^R_{t+1}, \{y_t^l - v^l + X^l_{t+1}\}_{l=1}^k)]\notag \\
&=\mathbb{E}[V_{t+1}(H^R_{t+1}, y_t^1 - v^1 + X^1_{t+1}, \ldots \notag \\
&\ldots, y_t^{i} - v^{i} + X^{i}_{t+1}, \ldots, \notag \\
&\ldots, y_t^{j} - v^{j} + X^{j}_{t+1}, \ldots,  y_t^{k} - v^k + X^{k}_{t+1})]\notag \\
&\le \mathbb{E}[V_{t+1}(H^R_{t+1}, y_t^1 - v^1 + X^1_{t+1}, \ldots \notag \\
&\ldots, y_t^{i} - (v^{i}-1) + X^{i}_{t+1}, \ldots, \notag \\
&\ldots, y_t^{j} - (v^{j}+1) + X^{j}_{t+1}, \ldots,  y_t^{k} - v^k + X^{k}_{t+1})]\notag \\
&=\mathbb{E}[V_{t+1}(H^R_{t+1}, \{y_t^l - a^l + X^l_{t+1}\}_{l=1}^k)].\notag 
\end{align}
Therefore the objective value  in (\ref{eq:v*})  can only improve after applying the transformation $\mathcal{T}(\cdot)$. \\

\textbf{Step 4:} Starting with any $v^{1:k} \neq v^{*1:k}$ in $\mathcal{V}(u^{1:k}, y_t^{1:k})$, we can keep applying transformation $\mathcal{T}(\cdot)$ to construct new vectors in $\mathcal{V}(u^{1:k}, y_t^{1:k})$ that result in an objective value at least as large as that under $v^{1:k}$. This is conducted in the following while-loop:
\begin{algorithmic}[1]
		\While{$v^{1:k} \neq v^{*1:k}$}
			\State $v^{1:k} \longleftarrow \mathcal{T}(v^{1:k})$ 
		\EndWhile
\State \textbf{return} $\; v^{1:k}$
  \end{algorithmic}
  The above while-loop will terminate in  finite number of steps with $v^{1:k} = v^{*1:k}$ at termination. Thus, the objective value  under  $v^{*1:k}$ is at least as large as that under any $v^{1:k} \in \mathcal{V}(u^{1:k}, y_t^{1:k})$. Thus, $v^{*1:k}$ is optimal.

\section{Proof of Theorem \ref{thm:opt-mech}} \label{sec:opt-mech-pf}
Suppose $n_t$ consumers arrive at time $t$ and let $(r,j)$ denote the  type reported by the $i$th consumer arriving at time $t$. Assuming that all other consumers report their types truthfully, let $Q^{*i}_t(r, j, n_t)$ and $P^{*i}_t(r, j, n_t)$ denote  the expected allocation and payment (see (\ref{eq:interim})-(\ref{eq:P})), respectively, for this consumer under the
mechanism $(q^*_{1:T}, p^*_{1:T})$, when it reports the pair $(r,j)$.

\textit{Bayesian Incentive Compatibility and Individual Rationality:} Because of Lemma \ref{lem:BICIR}, we can establish that $(q^*_{1:T}, p^*_{1:T})$  is Bayesian incentive compatible and individually rational by showing that the following conditions hold true:
\bromalist
\item $Q^{*i}_t(r, j, n_t)$ is non-decreasing in $r$ for all $i, t$.
\item $Q^{*i}_t(r, j, n_t)$ is non-decreasing in $j$ for all $i, t$.
\item   $P^{*i}_t(\theta^{\text{min}}, j, n_t) = 0$ for all $j,  n_t,  t,  i$.
\item $\theta^{\text{min}} \; Q^{*i}_t(\theta^{\text{min}}, j, n_t) = 0$ for all $j, n_t, t, i$.
\item $P^{*i}_t(r, j, n_t)$   is of  the form given in  (\ref{eq:interimP}) for all $i, t$.
\end{list}

We establish these conditions below:

(i) In order to establish that $Q^{*i}_t(r, j, \cdot)$ is non-decreasing in $r$, it suffices to show that $\sum\limits_{l \le j} q_t^{*i,l}(h^{-i,R}_{t}, (r, j), y_t^{1:k})$ is non-decreasing in $r$, where $h^{-i,R}_t$  denotes the set of  reports from all consumers other than $i$. Given $h^{-i,R}_t, y^{1:k}_t$  consider two information states $\ubar{s}_t := (h^{-i,R}_t, (\ubar{r}, j), y^{1:k}_t)$ and $\bar{s}_t := (h^{-i,R}_t, (\bar{r}, j), y^{1:k}_t)$,  where $\bar{r} > \ubar{r}$. That is, consumer $i$ has types $(\ubar{r}, j)$ and $(\bar{r}, j)$ under $\ubar{s}_t$ and $\bar{s}_t$, respectively. We now want to show that
\begin{align}\label{eq:monr}
\sum\limits_{l \le j} q^{*i,l}_t(\bar{s}_t) \ge \sum\limits_{l \le j} q^{*i,l}_t(\ubar{s}_t).
\end{align}

Clearly if  $\sum\limits_{l \le j} q^{*i,l}_t(\bar{s}_t)= 1$, (\ref{eq:monr}) holds true. 
Let's consider the case where $\sum\limits_{l \le j} q^{*i,l}_t(\bar{s}_t)= 0$. We need to argue that in this case $\sum\limits_{l \le j} q^{*i,l}_t(\ubar{s}_t) = 0$.
Let $u^{*1:k}$ denote the optimal $u^{1:k}$ vector obtained from solving the dynamic program in (\ref{eq:DPuv*t}) under the information state  $\bar{s}_t$. Since consumer $i$ does not get served under $u^{*1:k}$ (recall that $\sum\limits_{l \le j} q^{*i,l}_t(\bar{s}_t)= 0$), it can be shown that $u^{*1:k}$ is optimal under $\ubar{s}_t$ as well and that consumer $i$ will not get served under the information state $\ubar{s}_t$. In other words, $\sum\limits_{l \le j} q^{*i,l}_t(\ubar{s}_t) = 0$ and (\ref{eq:monr}) is true.

(ii) In order to establish that $Q^{*i}_t(r, j, \cdot)$ is non-decreasing in $j$, it suffices to prove that $\sum\limits_{l \le j} q_t^{*i,l}(h^{-i,R}_{t}, (r, j), y_{t}^{1:k})$ is non-decreasing in $j$.  We will use the following proposition in our proof.

\begin{prop}\label{prop:wmon}
Let $u^{*1:k}$  denote the optimal vector that results from solving the dynamic program  in (\ref{eq:DPuv*t}) under the information state  $s_t = (h^R_t, y_t^{1:k})$. Consider two flexibility levels $j$ and $j'$ with $j < j'$. Then, every consumer with flexibility level $j'$ and virtual valuation greater than $w_t^{u^{*j},j}$ gets served under $(q^*_{1:T}, p^*_{1:T})$.\footnote{If $u^{*j}=0$, then $w_t^{u^{*j},j} := \infty$.}
\begin{proof}
Suppose the proposition is not true.
Define the vector $\hat{u}^{1:k}$ as follows: $\hat{u}^{j} = u^{*j} - 1, \hat{u}^{j'} = u^{*j'} + 1$ and $\hat{u}^l = u^{*l}$ for all $l \neq j, j'$. Clearly $\hat{u}^{1:k} \in \mathcal{U}(h_t^{R}, y_{t}^{1:k})$. Consider the  expression of the value function  given in  (\ref{eq:DPuvt}) (which is equivalent to the definition in (\ref{eq:DPuv*t})). It is straightforward to verify that the first term in (\ref{eq:DPuvt}), i.e., $\sum\limits_{j=1}^{k} \sum\limits_{i=1}^{u^j} w_t^{i,j}$ would be strictly larger under the vector $\hat{u}^{1:k}$ compared to that under $u^{*1:k}$. Moreover, since $\mathcal{V}(u^{*1:k}, y_t^{1:k}) \subseteq \mathcal{V}(\hat{u}^{1:k}, y_t^{1:k})$ (see (\ref{eq:Vset})), the second term in (\ref{eq:DPuvt}) (i.e., the inner maximization over $v^{1:k}$ vector)  cannot decrease under the vector $\hat{u}^{1:k}$ compared to that under $u^{*1:k}$.  Therefore, the objective  in (\ref{eq:DPuvt}) (equivalently, (\ref{eq:DPuv*t}))  strictly improves under the  vector $\hat{u}^{1:k}$ compared to that evaluated at $u^{*1:k}$, which contradicts the optimality of $u^{*1:k}$.  This completes the proof.  
\end{proof}
\end{prop}

Given $h^{-i,R}_t, y^{1:k}_t$  consider two information states $\ubar{s}_t := (h^{-i,R}_t, (r,\ubar{c}), y^{1:k}_t)$ and $\bar{s}_t := (h^{-i,R}_t, (r,\bar{c}), y^{1:k}_t)$,  where $\bar{c}, \ubar{c} \in \{1, \ldots, k\}, \bar{c} > \ubar{c}$. Thus, we need to show that:
\begin{align}
&\sum\limits_{l \le \bar{c}} q_{t}^{*i,l}(\bar{s}_t) \ge \sum\limits_{l \le \ubar{c}} q_{t}^{*i,l}(\ubar{s}_t). \label{eq:xiMonb}
\end{align}
Clearly, if $\sum\limits_{l \le \bar{c}} q_{t}^{*i,l}(\bar{s}_t) = 1$, (\ref{eq:xiMonb}) holds true. 
Let's consider the case where $\sum\limits_{l \le \bar{c}} q_{t}^{*i,l}(\bar{s}_t) = 0$. We need to argue that in this case $\sum\limits_{l \le \ubar{c}} q_{t}^{*i,l}(\ubar{s}_t) = 0$. Let $u^{*1:k}$ denote the optimal $u^{1:k}$ vector obtained from solving the dynamic program  in (\ref{eq:DPuv*t}) under the information state $\bar{s}_t$. 
Let $n^j_t(s_t)$ denote the number of consumers with flexibility level $j$ that arrive at time $t$ under the information state $s_t$.  Because consumer $i$ with flexibility level $\bar{c}$ does not get a good under $\bar{s}_t$  (recall that $\sum\limits_{l \le \bar{c}} q_{t}^{*i,l}(\bar{s}_t) = 0$), it clearly means that $u^{*\bar{c}} \le n^{\bar{c}}_t(\bar{s}_t) - 1 = n^{\bar{c}}_t(\ubar{s}_t)$. Therefore, indeed $u^{*1:k} \in \mathcal{U}(\ubar{s}_t)$ (see (\ref{eq:Uset})). 
We now want to show that $u_t^{*1:k}$ is also optimal under  $\ubar{s}_t$.

Consider the following sequence of implications:
\begin{enumerate}[(a)]
\item Since consumer $i$ does not get served under the information state $\bar{s}_t$ (recall that $\sum\limits_{l \le \bar{c}} q^{*i,l}_t(\bar{s}_t)= 0$), it follows from Proposition \ref{prop:wmon}  that its virtual valuation must be no greater than the virtual valuations of the  consumers that are served from  flexibility levels \textit{lower} than $\bar{c}$; in particular $w_t(r,\bar{c}) \le w_t^{u^{*\ubar{c}}, \ubar{c}}$.  
\item From Assumption \ref{assum:mhrc} we know that $w_t(r,\ubar{c}) < w_t(r,\bar{c})$. Therefore, (a) implies that $w_t(r,\ubar{c}) < w_t^{u^{*\ubar{c}}, \ubar{c}}$.
\item $\sum\limits_{l \le \bar{c}} q^{*i,l}_t(\bar{s}_t)= 0$ combined with (b) implies that the vector $u^{*1:k}$ results in the exact same objective value in (\ref{eq:DPuv*t}) under both information states $\bar{s}_t$ and $\ubar{s}_t$.
\item Since   $\ubar{c}<\bar{c}$ and $w_t(r,\ubar{c}) < w_t(r,\bar{c})$, it is straightforward to show that under the information state $\ubar{s}_t$, the value function  in (\ref{eq:DPuv*t})  is upper bounded by that under $\bar{s}_t$, i.e., $V_t(\ubar{s}_t)  \le V_t(\bar{s}_t)$.
\item Items (c) and (d) combined, imply that $u^{*1:k}$ is optimal under $\ubar{s}_t$ as well. 
\end{enumerate}
Items (e) and (b) above imply that consumer $i$ with type $(r,\ubar{c})$ does not get served under the information state $\ubar{s}_t$, i.e., $\sum\limits_{l \le \ubar{c}} q^{*i,l}_t(\ubar{s}_t) = 0$. Thus, (\ref{eq:xiMonb}) is true.

(iii)-(v): To establish conditions (iii)-(v) consider  the payment form given below:
\begin{align}
&\rho_t^{*i}(h^{-i,R}_t, (r,j), y_t^{1:k}) = r \sum\limits_{j' \le j} q_t^{*i,j'}(h^{-i,R}_t, (r,j), y_t^{1:k})  \notag \\
&- \int\limits_{\theta^{\text{min}}}^{r} \Big(\sum\limits_{j' \le j} q_t^{*i,j'}(h^{-i,R}_t, (\alpha, j), y_t^{1:k}) \Big) \; d\alpha. \label{eq:rho*}
\end{align}
We first argue that monotonicity of $q^*_t(\cdot)$ as established above  (condition (i)), implies that the payment form in  (\ref{eq:rho*}) is equivalent to the one given in (\ref{eq:p*}). Then we show that this equivalent payment form in  (\ref{eq:rho*}), indeed satisfies conditions (iii)-(v).

If consumer $i$ with the  report $(r, j)$ does not get a good (i.e., $\sum\limits_{j' \le j} q_t^{*i,j'}(h^{-i,R}_t, (r,j), y_t^{1:k}) = 0$), then the monotonicity of $q^*_t$  implies that the integral in (\ref{eq:rho*}) is also 0. Hence, in this case consumer $i$ pays nothing. On the other hand, if  consumer $i$ gets a good (i.e., $\sum\limits_{j' \le j} q_t^{*i,j'}(h^{-i,R}_t, (r,j), y_t^{1:k}) = 1$), then the definition of $\bar{\theta}^{i,j}_t$ (see (\ref{eq:thetabarijt})) implies that the integral in (\ref{eq:rho*}) is $(r - \bar{\theta}^{i,j}_t)$. Hence, in this case consumer $i$ pays $\bar{\theta}^{i,j}_t$. Thus, the payment in (\ref{eq:rho*}) is identical to the payment in (\ref{eq:p*}).

We now argue that the equivalent expression for $p^*_t(\cdot)$ in (\ref{eq:rho*}) satisfies the conditions in (iii)-(v).

To see that condition (iii) above holds true, recall that from Assumption \ref{assum:mhrc} we have that $w_t(\theta^{\text{min}}, j) < 0$ for all $j, t$. Hence, it must be that $\sum\limits_{j' \le j} q_t^{*i,j'}(h^{-i,R}_{t}, (\theta^{\text{min}}, j), y_t^{1:k}) = 0$ for all $h^{-i,R}_{t}, y_t^{1:k}, j, i , t$. Otherwise $q^*_{1:T}$ will not maximize the objective in \eqref{revenue} whose solution is given by the dynamic program in (\ref{eq:DPVqtempty})-(\ref{eq:DPVqt}). Therefore, from (\ref{eq:rho*}) it follows that $\rho^{*i}_{t}(h^{-i,R}_{t}, (\theta^{\text{min}}, j), y_t^{1:k}) = 0$ for all $h^{-i,R}_{t}, y_t^{1:k}, j, i , t$. This  implies that $P^{*i}_t(\theta^{\text{min}}, j, n_t) = 0$ for all $j,  n_t,  t,  i$ which establishes condition (iii) above. Based on the same argument condition (iv) above holds true as well.

By taking the expectation of $\rho^{*i}_{t}(H^{-i,R}_{t}, (r,j), Y_t^{1:k})$ in (\ref{eq:rho*}) over $(H^{-i,R}_{t}, Y_t^{1:k})$, where $H^{-i,R}_{t} = H^R_t \setminus \{(\theta^i_t, b^i_t)\}$, it is easily established that the expected payment $P^{*i}_t(\cdot)$ satisfies (\ref{eq:interimP}) with $\theta^{\text{min}} \; Q^{*i}_t(\theta^{\text{min}}, j, n_t)  = 0$. Hence condition (v) above holds true.

The above arguments establish that the mechanism $(q^*_{1:T}, p^*_{1:T})$  is Bayesian incentive compatible and individually rational.

\textit{Expected-revenue maximization:}
The allocation functions $q^*_{1:T}$ constructed in Theorem \ref{thm:opt-mech} are  the optimal control strategy for the stochastic control problem in \eqref{revenue}. This is because  Lemma \ref{lem:infoState}-\ref{lem:v*recipe}  established that the dynamic program in (\ref{eq:DPuv*t}) is  equivalent to the one in (\ref{eq:DPqtempty})-(\ref{eq:DPqt}), which was formulated to address the control strategy optimization in \eqref{revenue}. Moreover, the payment functions $p^*_{1:T}$ defined in (\ref{eq:p*}) (which is equivalent to (\ref{eq:rho*})) satisfy (\ref{eq:expostp}).
Therefore, based on the results of Lemma \ref{lem:revEqBern}, the  mechanism $(q_{1:T}^*, p_{1:T}^*)$ is an expected revenue maximizing BIC and IR mechanism.

\bibliographystyle{IEEEtran}
\bibliography{REF}

\begin{thebibliography}{10}
\providecommand{\url}[1]{#1}
\csname url@samestyle\endcsname
\providecommand{\newblock}{\relax}
\providecommand{\bibinfo}[2]{#2}
\providecommand{\BIBentrySTDinterwordspacing}{\spaceskip=0pt\relax}
\providecommand{\BIBentryALTinterwordstretchfactor}{4}
\providecommand{\BIBentryALTinterwordspacing}{\spaceskip=\fontdimen2\font plus
\BIBentryALTinterwordstretchfactor\fontdimen3\font minus
  \fontdimen4\font\relax}
\providecommand{\BIBforeignlanguage}[2]{{%
\expandafter\ifx\csname l@#1\endcsname\relax
\typeout{** WARNING: IEEEtran.bst: No hyphenation pattern has been}%
\typeout{** loaded for the language `#1'. Using the pattern for}%
\typeout{** the default language instead.}%
\else
\language=\csname l@#1\endcsname
\fi
#2}}
\providecommand{\BIBdecl}{\relax}
\BIBdecl

\bibitem{agmon2013deconstructing}
O.~Agmon Ben-Yehuda, M.~Ben-Yehuda, A.~Schuster, and D.~Tsafrir,
  ``Deconstructing amazon ec2 spot instance pricing,'' \emph{ACM Transactions
  on Economics and Computation}, vol.~1, no.~3, p.~16, 2013.

\bibitem{rogers2012delivering}
A.~Rogers, S.~D. Ramchurn, and N.~R. Jennings, ``Delivering the smart grid:
  Challenges for autonomous agents and multi-agent systems research,'' in
  \emph{Twenty-Sixth AAAI Conference on Artificial Intelligence}, 2012.

\bibitem{hossain2009dynamic}
E.~Hossain, D.~Niyato, and Z.~Han, \emph{Dynamic spectrum access and management
  in cognitive radio networks}.\hskip 1em plus 0.5em minus 0.4em\relax
  Cambridge university press, 2009.

\bibitem{bergemann2019dynamic}
D.~Bergemann and J.~V{\"a}lim{\"a}ki, ``Dynamic mechanism design: An
  introduction,'' \emph{Journal of Economic Literature}, vol.~57, no.~2, pp.
  235--74, 2019.

\bibitem{bergemann2010dynamic}
D.~Bergemann and M.~Said, ``Dynamic auctions: A survey,'' 2010.

\bibitem{nisan2007algorithmic}
N.~Nisan, T.~Roughgarden, E.~Tardos, and V.~V. Vazirani, \emph{Algorithmic game
  theory}.\hskip 1em plus 0.5em minus 0.4em\relax Cambridge University Press
  Cambridge, 2007, vol.~1.

\bibitem{vulcano2002optimal}
G.~Vulcano, G.~Van~Ryzin, and C.~Maglaras, ``Optimal dynamic auctions for
  revenue management,'' \emph{Management Science}, vol.~48, no.~11, pp.
  1388--1407, 2002.

\bibitem{pai2013optimal}
M.~M. Pai and R.~Vohra, ``Optimal dynamic auctions and simple index rules,''
  \emph{Mathematics of Operations Research}, vol.~38, no.~4, pp. 682--697,
  2013.

\bibitem{lavi2004competitive}
R.~Lavi and N.~Nisan, ``Competitive analysis of incentive compatible on-line
  auctions,'' \emph{Theoretical Computer Science}, vol. 310, no. 1-3, pp.
  159--180, 2004.

\bibitem{gallien2006dynamic}
J.~Gallien, ``Dynamic mechanism design for online commerce,'' \emph{Operations
  Research}, vol.~54, no.~2, pp. 291--310, 2006.

\bibitem{gershkov2017revenue}
A.~Gershkov, B.~Moldovanu, and P.~Strack, ``Revenue-maximizing mechanisms with
  strategic customers and unknown, markovian demand,'' \emph{Management
  Science}, vol.~64, no.~5, pp. 2031--2046, 2017.

\bibitem{said2012auctions}
M.~Said, ``Auctions with dynamic populations: Efficiency and revenue
  maximization,'' \emph{Journal of Economic Theory}, vol. 147, no.~6, pp.
  2419--2438, 2012.

\bibitem{gershkov2009dynamic}
A.~Gershkov and B.~Moldovanu, ``Dynamic revenue maximization with heterogeneous
  objects: A mechanism design approach,'' \emph{American economic Journal:
  microeconomics}, vol.~1, no.~2, pp. 168--98, 2009.

\bibitem{mierendorff2016optimal}
K.~Mierendorff, ``Optimal dynamic mechanism design with deadlines,''
  \emph{Journal of Economic Theory}, vol. 161, pp. 190--222, 2016.

\bibitem{borgers2015introduction}
T.~Borgers, R.~Strausz, and D.~Krahmer, \emph{An introduction to the theory of
  mechanism design}.\hskip 1em plus 0.5em minus 0.4em\relax Oxford University
  Press, USA, 2015.

\bibitem{navabi2018optimal}
S.~Navabi and A.~Nayyar, ``Optimal auction design for flexible consumers,''
  \emph{IEEE Transactions on Control of Network Systems}, 2018.

\bibitem{puterman2014markov}
M.~L. Puterman, \emph{Markov decision processes: discrete stochastic dynamic
  programming}.\hskip 1em plus 0.5em minus 0.4em\relax John Wiley \& Sons,
  2014.

\end{thebibliography}
 
\end{document}